\documentclass[aps, pra,superscriptaddress, twocolumn,showkey]{revtex4-1}
\usepackage{comment}
\usepackage{enumerate}
\usepackage{amssymb}
\usepackage{amsmath}
\usepackage{graphicx}
\usepackage[usenames,dvipsnames]{color}

\usepackage[colorlinks,bookmarks=false,citecolor=NavyBlue,linkcolor=OliveGreen,urlcolor=blue]{hyperref}

\def\doi{http://dx.doi.org/}

\begin{document}
\title{Spreading of correlations in Markovian open quantum systems}
\author{Vincenzo Alba}
\address{Institute  for  Theoretical  Physics, Universiteit van Amsterdam,
Science Park 904, Postbus 94485, 1098 XH Amsterdam,  The  Netherlands}

\author{Federico Carollo}
\address{Institut f\"{u}r Theoretische Physik, Universit\"{a}t T\"{u}bingen, Auf der Morgenstelle 14, 72076 T\"{u}bingen, Germany
}

\begin{abstract}

Understanding the spreading of quantum correlations in out-of-equilibrium many-body systems is one of the major challenges in physics. For {\it isolated} systems, a hydrodynamic theory explains the origin and spreading of entanglement via the propagation of quasi-particle pairs. However, when systems interact with their surrounding much less has been established. Here we show that the quasi-particle picture remains valid for open quantum systems: while information is still spread by quasiparticles, the environment modifies their correlation and introduces incoherent and mixing effects. For free fermions with gain/loss dissipation we provide
formulae fully describing incoherent and quasiparticle 
contributions in 
the spreading of entropy and mutual 
information. Importantly, the latter is not affected by 
incoherent correlations. The mutual information is exponentially damped at short times and eventually vanishes signalling the onset of a classical limit. The behaviour of the logarithmic negativity is similar and this scenario 
is common to other dissipations. For weak dissipation, the presence of 
quasiparticles underlies remarkable scaling behaviors.

\end{abstract}

\maketitle

\paragraph*{Introduction.--}

Recent years have witnessed tremendous breakthroughs in understanding 
the origin and spreading of entanglement in out-of-equilibrium 
many-body systems. In particular, a well-established hydrodynamic picture, 
predicated on the existence of stable quasiparticles, 
allows for the description of the entanglement dynamics in integrable 
systems~\cite{cc-05,fagotti-2008,alba-2016}. 
The unique ability of these systems to transfer quantum correlations
can be potentially leveraged in quantum technologies~\cite{sougato-2007} and their implementations in quantum computers are now 
within reach~\cite{smith-2019}. 
Unfortunately, in realistic settings, a coupling with an environment is unavoidable and typically induces a quick decay of quantum coherence. 

This loss of quantum information is also the key of the 
black-hole information paradox~\cite{hawking-1,hawking-2,hayden-2007}. 
Black holes can evaporate by emitting Hawking radiation in 
a perfectly mixed state. 
Thus, one is left with the conundrum that an initially 
pure state (the black hole) gets transformed 
in a mixed one (radiation) after the black-hole evaporates. 
As it was pointed out by Page~\cite{page-1,page-2}, 
the paradox is apparent after the black hole is half 
evaporated at $t_\mathrm{Page}\approx t_\mathrm{evap}/2$. 
Interestingly, operational schemes to solve this paradox 
are based on decoherence~\cite{bao-2018,bao-2019}.
Their virtue is that even a small decoherence dramatically reduces 
the mutual information between the inside and the 
outside of the black-hole~\cite{agarwal-2019}, delaying the onset of the  
paradox to times when the black hole is 
nearly evaporated, and the semiclassical treatment~\cite{hawking-1} 
might break down. 

This demonstrates how understanding the fate of entaglement in open quantum systems will have immediate impact on very different branches of physics. 
Still, modelling quantum information loss, or even sieving genuine quantum 
coherence from classical correlations, is considered a daunting task. 
Here, for the first time, we show that these aspects can be understood within the hydrodynamic theory and encapsulated in properties of quasi-particles.  

We adopt the framework of Markovian open quantum dynamics~\cite{petruccione}. In these settings (weak system-environment coupling and absence of memory effects), the time-evolution of any initial density matrix $\rho_0$ is generated by the Lindblad master equation $\dot{\rho}_t=\mathcal{L}[\rho_t]$ 
with the Liouvillian ${\mathcal L}$ being~\cite{petruccione}
\begin{equation}
\label{l-eq}
{\mathcal L}[\rho]=-i[H,\rho]+
\sum_{\mu}\Big(L_\mu\rho L^\dagger_\mu-
\frac{1}{2}\{L^\dagger_\mu L_\mu,\rho\}\Big). 
\end{equation}
Here  $-i[H,\rho]$ describes the coherent evolution 
under a many-body Hamiltonian $H$, whereas Lindblad operators $L_\mu$ effectively account for the presence of an environment. 

%
\begin{figure}[t]
\includegraphics[width=0.5\textwidth]{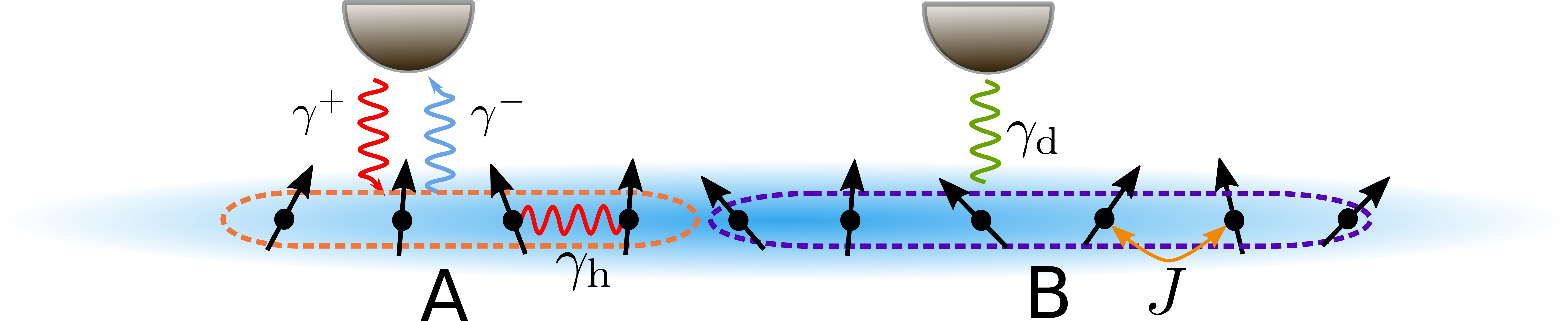}
\caption{Sketch of a fermionic open quantum chain. We consider different dissipative 
 effects. Rates $\gamma^\pm$ are related to fermion creation and annihilation. $\gamma_h$ 
 and $\gamma_d$ are, instead, rates for incoherent hopping of fermions and dephasing, 
 respectively. Here the coefficient $J$ represents the amplitude of coherent 
 hopping. The chain is bipartite as $A\cup B$. 
}
\label{fig0:chain}
\end{figure}
%

We consider a fermionic chain with $L$ sites 
(see Fig.~\ref{fig0:chain}) and denote standard creation and annihilation operators 
as $c_m^\dagger,c_m$ for each site. We focus on the XX chain Hamiltonian $H=\sum_{m,n=1}^{L}h_{mn}c^\dagger_m c_n+\mathrm{h.c.}$ with $h_{nm}=J\delta_{|n-m|,1}$. We discuss several sources of dissipation, namely gain/loss, 
i.e., creation and annihilation of fermions in all sites, 
incoherent hopping and dephasing. Gain/loss is 
described by $L_m^+=\sqrt{\gamma^{\scriptscriptstyle +}} c^\dagger_m$ and 
$L^-_m=\sqrt{\gamma^{\scriptscriptstyle -}} c_m$. 
A similar Lindbladian has been considered very recently in 
Ref.~\onlinecite{maity-2020}. 
Incoherent hopping~\cite{eisler-2011} corresponds to 
$L_{L,m}=\sqrt{\gamma_h}c^\dagger_m c_{m+1}$  and $L_{R,m}=L_{L,m}^\dagger$ describing jumps of particles on the left or on the right. Finally, dephasing is modelled by 
$L_{d,m}=\sqrt{\gamma_d}c^\dagger_m c_m$. We study the 
dynamics ensuing from the N\'eel state $|N\rangle\equiv\left|\uparrow
\downarrow\uparrow\cdots\right\rangle$, although any product state 
can be treated analogously. 

To quantify information spreading we divide the 
system into two parts, as $A\cup B$ (see Fig.~\ref{fig0:chain}). 
From the reduced density matrix $\rho_{A/B}$ we define the R\'enyi entropies  
$S^{(\alpha)}_{A/B}=1/(1-\alpha)\ln\mathrm{Tr}\rho_{A/B}^\alpha, 
\quad\mathrm{with}\, \alpha\in\mathbb{R}$. The von Neumann entropy is obtained by taking $\alpha\to1$ as 
$S_{A/B}\equiv -\mathrm{Tr}\rho_{A/B}\ln\rho_{A/B}$. The mutual 
information $I_A^{\scriptscriptstyle(\alpha)}=S_A^{\scriptscriptstyle(\alpha)}+
S_{A}^{\scriptscriptstyle(\alpha)}-
S_{A\cup B}^{\scriptscriptstyle(\alpha)}$ is thus a measure of the total correlation 
between a $A$ and $B$. The logarithmic negativity~\cite{lee-2000,vidal-2002,eisert-1999,plenio-2005,hannu-2008,mrpr-09,calabrese-2012,cct-neg-long,shapourian-2016,ssr-16,coser-2017} is instead a genuine 
measure of the entanglement. 
This is obtained from the partial transpose 
$\rho_A^{T}$, with $\langle\varphi_i,\bar \varphi_j|
\rho_A^T|\varphi_i',\bar\varphi_j'\rangle=
\langle\varphi_i,\bar \varphi_j'|\rho_A^T|\varphi_i',\bar\varphi_j\rangle$,  
where $|\varphi_i\rangle$ and $|\bar\varphi_i\rangle$ are two orthonormal 
bases for $A$ and $B$. For free fermions, $\rho_A^T=(e^{-i\frac{\pi}{4}}O_++e^{i\frac{\pi}{4}}O_-)/\sqrt{2}$, 
with $O_\pm$ two gaussian operators~\cite{eisler-2014a} . 
Thus, $\rho_A^{T}$ is not gaussian and the computation of the logarithmic negativity is difficult even for free fermions.  
Recently, an alternative negativity, ${\cal E}\equiv\ln\mathrm{Tr}\sqrt{O_+O_-}$, has been introduced~\cite{ssr-16}.
${\cal E}$ is an entanglement monotone under 
local operations and classical communication preserving the local fermion-number parity~\cite{shapourian-2018a} and can be easily computed (see Appendix~\ref{sec:negat}).

In out-of-equilibrium integrable systems after a 
 {\it quantum quench}~\cite{calabrese-2016,essler-2016,
vidmar-2016,caux-2016}, 
the quasiparticle picture~\cite{cc-05,fagotti-2008,alba-2016,alba-2018} 
allows one to describe the dynamics of the von Neumann entropy~\cite{alba-2016}, 
the steady-state R\'enyi entropies~\cite{renyi,renyi-1,mestyan-2018,alba-2019}, and the mutual 
information~\cite{alba-2018,alba-2019b}, also for quenches 
from inhomogeneous initial states~\cite{alba-2018a,bertini-2018,
alba-2019a,mestyan-2020}. 
Remarkably, it has been shown that 
${\cal E}=I^{\scriptscriptstyle (1/2)}_A/2$~\cite{alba-epl}, then verified in  
holographic calculations~\cite{flam-2020-1}, and observed 
in systems with defects~\cite{gruber-2020}. 
In the quasiparticle picture the initial state acts as  a source of 
pairs (or multiplets~\cite{bertini-2018a,bastianello-2018,bastianello-2020}) 
of entangled quasiparticles. As 
they propagate, they entangle larger and larger portions of 
the system. The entanglement entropy $S(t)$ is proportional 
to the number of entangled pairs  that at time $t$ are shared 
between $A$ and $B$.
%
\begin{figure}[t]
\includegraphics[width=0.45\textwidth]{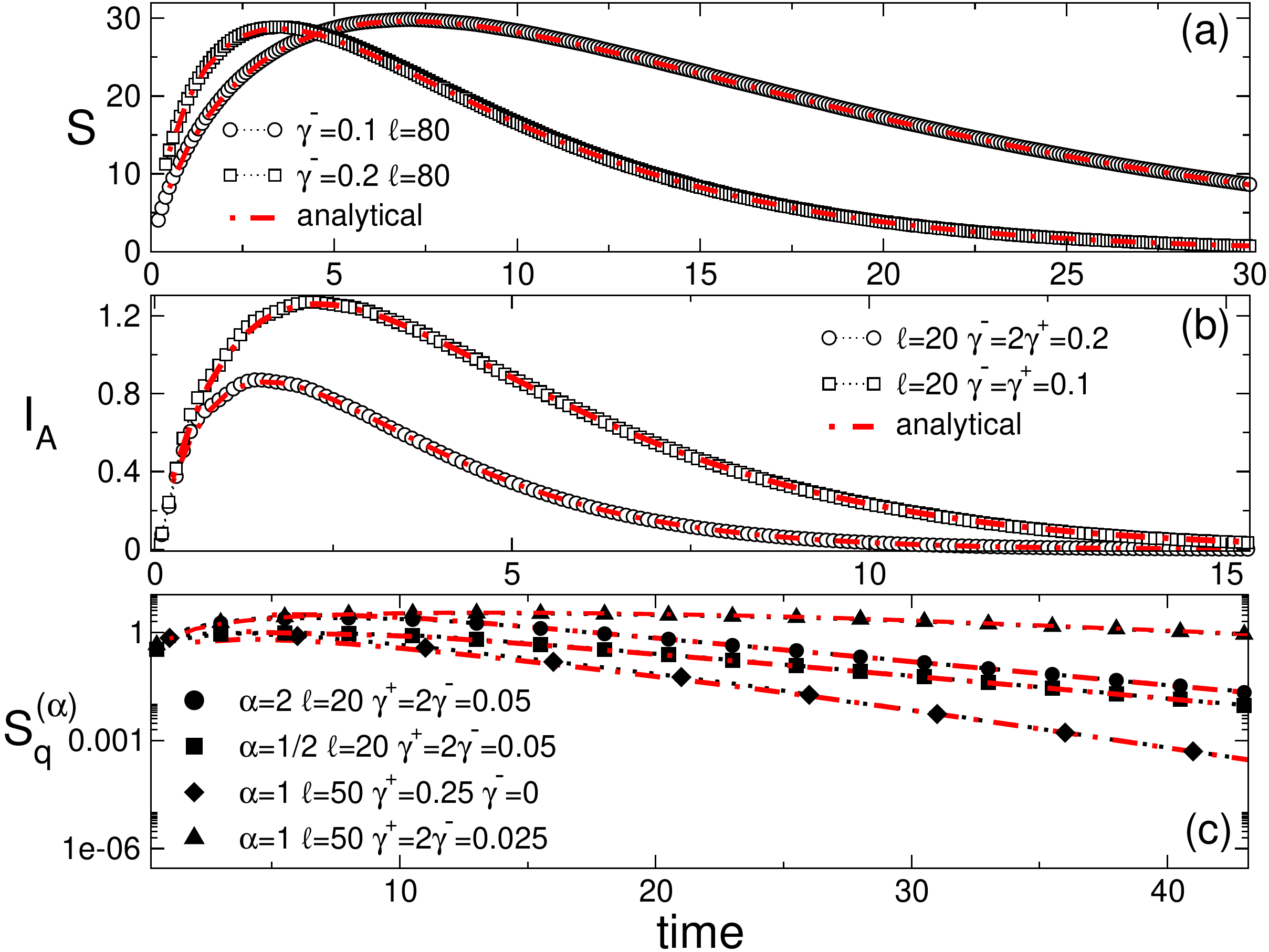}
\caption{Effect of gain/loss dissipation in the $XX$ chain after 
 the N\'eel quench. $\gamma^\pm$ are dissipation rates. (a) Dynamics 
 of the entropy $S$ for $\gamma^+=0$. $\ell$ is the subsytem size.  
 Symbols are exact numerical data. Dashed-dotted lines denote~\eqref{main-formula}. 
 (b) Dynamics of $I_A$. (c) Dynamics of $S_\mathrm{q}$ (cf.~\eqref{boxed}). Now the 
 dashed dotted lines denote the large $t$ results (c.f.~\eqref{boxed-1}
 \eqref{renyi-theo}). 
}
\label{fig:quantum-entropy}
\end{figure}
%
For generic integrable systems, both interacting and free, the 
local steady-state physics is described by a Generalized Gibbs 
Ensemble~\cite{calabrese-2016,essler-2016,vidmar-2016,caux-2016} (GGE). 
Remarkably, the entanglement content of the quasiparticles is 
the GGE thermodynamic entropy, whereas the 
quasiparticles velocity is calculated from the excitations 
above the GGE macrostate~\cite{alba-2016}. 
%
\begin{figure*}[t]
\includegraphics[width=0.9\textwidth]{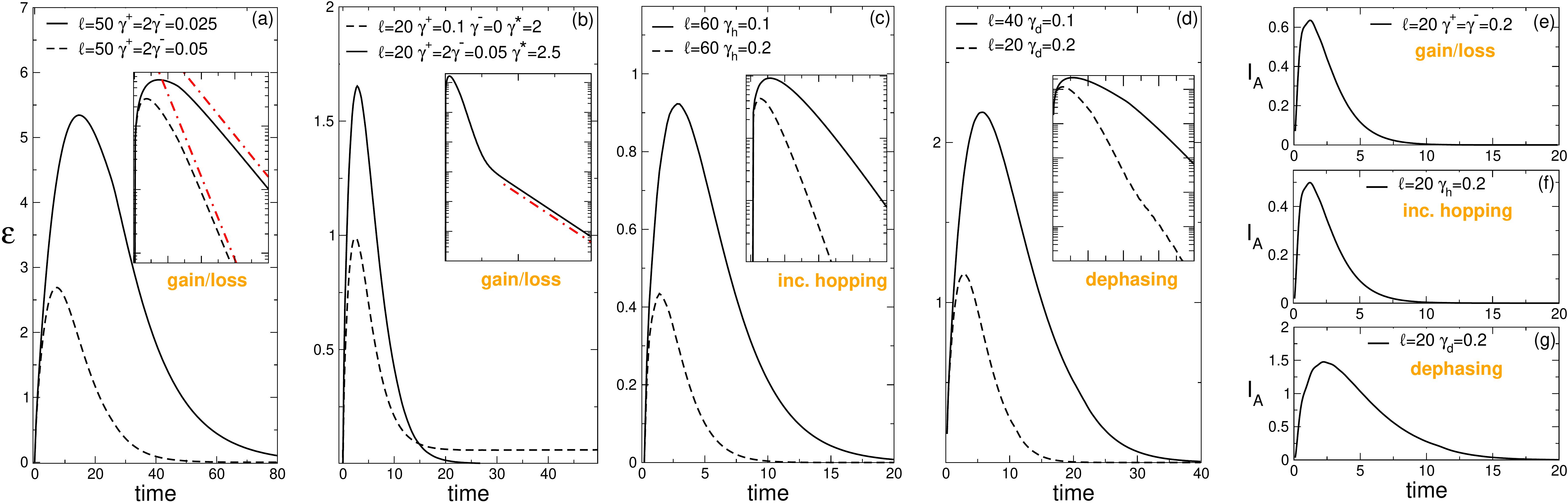}
\caption{Entanglement dynamics after the 
 N\'eel quench in the XX chain with dissipation: 
 diagonal gain/loss with rates $\gamma^\pm$ (a), 
 non-diagonal gain/loss with rates $\gamma^\pm,\gamma^*$ (b), 
 incoherent hopping with rate $\gamma_h$ (c), and dephasing with rate 
 $\gamma_d$ (d). Panels (a-d) show ${\cal E}$. In all cases entanglement 
 is created only up to $t\approx1/\gamma$ with $\gamma=\gamma^\pm,\gamma_h,\gamma_d$. 
 For diagonal gain/loss one has ${\cal E}\propto \exp({-2(\gamma^++\gamma^-)}t)$ 
 (dashed dotted line in the inset in (a)). In (b) the dashed-dotted line is  
 ${\cal E}\propto\exp(-2(\gamma^-+\gamma^+)(\gamma-2)t)$. Note that for 
 $\gamma=2$ the generator is gapless. Dissipation is off-diagonal 
 resulting in a finite steady state value of ${\cal E}$. For incoherent hopping 
 and dephasing ${\cal E}$ decays at $t\ll\ell$. For $t\gg \ell$, i.e., in 
 the steady state,  ${\cal E}$ attains a finite small value. (e-g) 
 Mutual information $I_A$ plotted as a function of time. 
}
\label{fig:neg-panel}
\end{figure*}
%
How does such scenario change in open quantum systems?
Here we show that for open quantum free-fermionic systems subject to linear~\cite{prosen-2008} 
diagonal dissipation the information dynamics is indeed  
captured by a modified quasiparticle picture. 
This allows for a complete analytical description of 
$S^{\scriptscriptstyle(\alpha)}$ and $I_A^{\scriptscriptstyle(\alpha)}$. 
Crucially, the environment affects 
the correlation content of the quasiparticles, and creates 
also classical contributions. 
Interestingly, the mutual information is only sensitive to 
the contribution from quasi-partices, and its dynamics closely reflects that of the negativity, despite not being a proper entanglement measure. For weak dissipation we 
reveal the remarkable scaling behavior 
\begin{equation}
\label{eq:scaling}
\gamma S^{(\alpha)}=f_\alpha(\gamma t,\gamma\ell), \quad 
\gamma{\cal E}=g(\gamma t,\gamma\ell),
\end{equation}
with $f_\alpha(x)$ and $g(x)$ two scaling functions, and $\gamma$ 
the relevant dissipation rate. Similar scaling 
persists for both incoherent hopping and dephasing, suggesting robustness of the quasiparticle picture to non-quadratic dissipative contributions. 
%
\begin{figure}[t]
\includegraphics[width=0.45\textwidth]{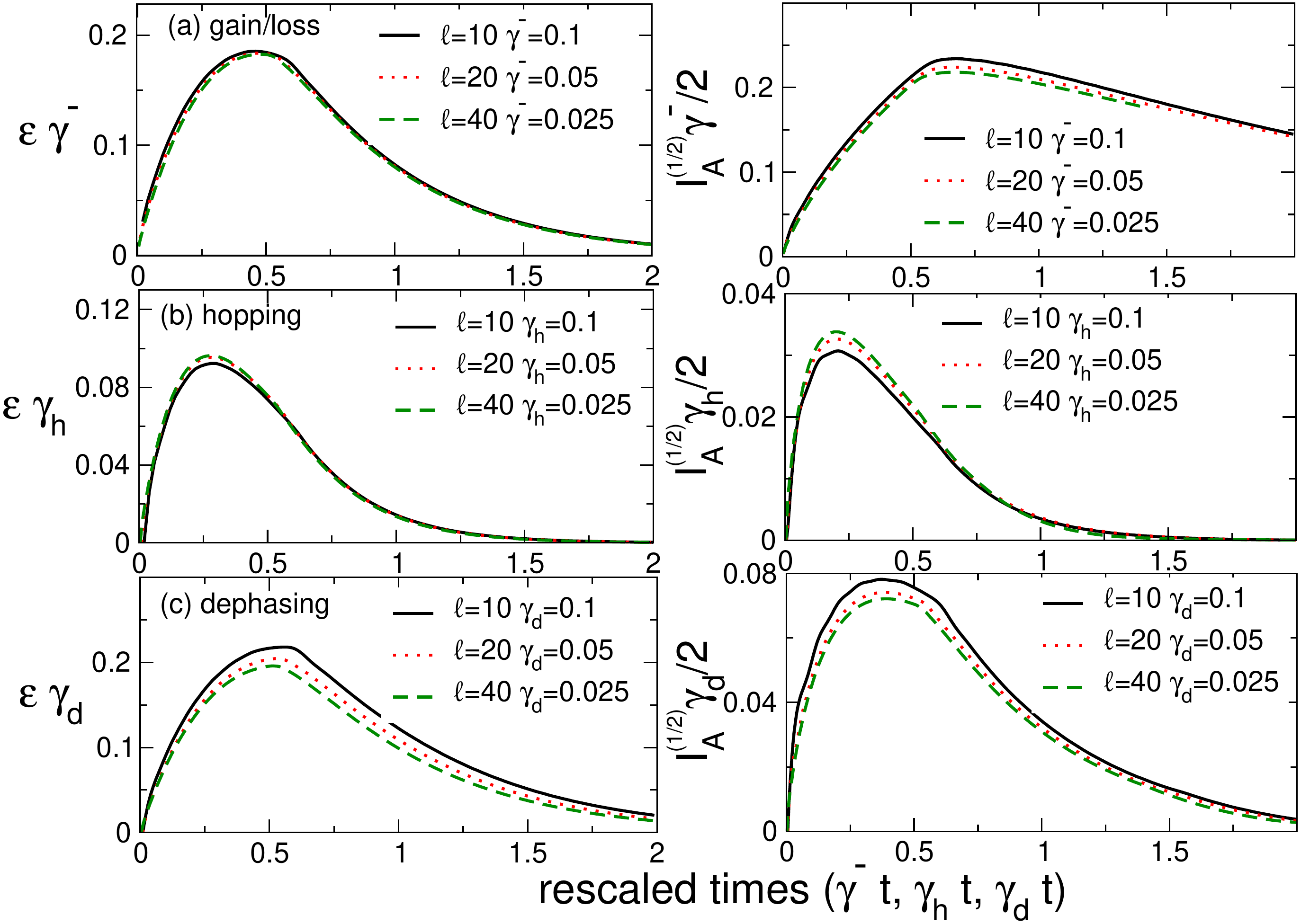}
\caption{ Scaling behavior of ${\cal E}$ and of $I^{\scriptscriptstyle 
 (1/2)}_A$ (left and right column, respectively). Different panels are for different types 
 of dissipation: gain/loss (a), incoherent hopping (b), and dephasing (c), 
 with dissipation rates $\gamma^+,\gamma_h,\gamma_d$. The scaling limit 
 is defined as $t,\ell,\to\infty$, $\gamma\to0$, with $t/\ell$ and 
 $\gamma\ell$, $\gamma t$ fixed. 
}
\label{fig:negscal}
\end{figure}
%
\paragraph*{Gain/loss dissipation.--} 
For gain/loss dissipation the Lindblad generator~\eqref{l-eq} 
is quadratic. The ensuing dynamics maps Gaussian states 
onto Gaussian states. The system 
is thus completely characterized by its two-point correlation functions. 
Assuming that at $t=0$ one has $\langle c_m c_n\rangle=0$ 
and $\langle c_m\rangle=0$, the relevant covariance matrix is  
$G_t\equiv\langle c^\dagger_m c_n\rangle_t$. 
The R\'enyi entropies are 
obtained from the eigenvalues $\lambda_i$ of $G_{t}$ restricted to $A$  
(cf.~\eqref{eig}) as~\cite{peschel-2009} 
$S^{(\alpha)}=1/(1-\alpha)\sum_i\ln[\lambda_i^\alpha+(1-\lambda_i)^\alpha]$. 
The time-evolved $G_t$ is given by
\begin{equation}
G_t=e^{t \Lambda}G_0e^{t \Lambda^\dagger} +\int_0^tdz 
\, e^{(t-z)\, \Lambda}\Gamma^+e^{(t-z)\, \Lambda^\dagger}\, ,
\label{CovMat}
\end{equation}
with $\Lambda=i h-1/2(\Gamma^++\Gamma^-)$, where $h$ is the Hamiltonian matrix, 
and $\Gamma^\pm_{mn}=\gamma^\pm\delta_{mn}$. 

It is useful to introduce the unitarily-evolved correlation matrix 
$\widetilde G_t=e^{i h t}G_0e^{-i h t}$. Since $\Gamma^\pm$ is diagonal, 
the eigenvalues $\lambda_i$ of $G_t$ are 
\begin{equation}
\label{eig}
\lambda_i=n_\infty(1-b(t))+\widetilde\lambda_i b(t),
\end{equation}
with $\widetilde\lambda_i$ being the eigenvalues of $\widetilde G_t$, and
\begin{equation}
\label{n-steady}
n_\infty\equiv\frac{\gamma^+}{\gamma^++\gamma^-},\quad b(t)\equiv e^{-(\gamma^++\gamma^-)t}. 
\end{equation}
Let us first consider a quench from the ferromagnet $|\mathrm{F}
\rangle\equiv\left|\downarrow\downarrow\cdots\right\rangle$, with homogeneous local fermionic occupation $n_t$. 
Physically,  $n_t$ satisfies the simple 
rate equation 
\begin{equation}
\label{rate}
d n_t=(1-n_t)\gamma^+ dt-\gamma^-n_t dt. 
\end{equation}
Eq.~\eqref{rate} reflects that the probability to create a 
fermion in the interval $[t,t+dt]$  is $(1-n_t)\gamma^+dt$, 
whereas a fermion is removed with probabilty $n_t\gamma^-dt$. 
The solution of~\eqref{rate} is straightforward as 
\begin{equation}
\label{rate-sol}
n_t=n_\infty\Big\{1-\Big[(1-n_0)-\frac{\gamma^-}{\gamma^+}n_0
\Big]e^{-(\gamma^++\gamma^-)t}\Big\}, 
\end{equation}
where $n_0$ is the initial occupation.   
Let us now consider the N\'eel state $|\mathrm{N}\rangle\equiv\left|
\uparrow\downarrow\uparrow\cdots\right\rangle$. 
Translational invariance is restored during the dynamics, 
and $n_t$ is obtained by replacing $n_0$ with its average 
$\langle n_0\rangle$ in Eq.~\eqref{rate-sol}. For the N\'eel state 
$\langle n_0\rangle=1/2$. The validity of Eq.~\eqref{rate} and~\eqref{rate-sol} 
is numerically verified in Appendix~\ref{sec:ploss-classical}. 

For open systems the R\'enyi  entropies contain 
both incoherent and coherent contributions, where with coherent we mean those attributed to  quasiparticles. 
As we shall see, a crucial role is played by the Yang-Yang entropies 
$S^{\scriptscriptstyle(\alpha)}_\mathrm{cl}$, defined from 
$n_t$ (cf.~\eqref{rate-sol}) as 
\begin{equation}
\label{renyi-yy}
S_{\mathrm{cl}}^{(\alpha)}=\frac{\ell}{1-\alpha}\ln(n_t^\alpha+(1-n_t)^\alpha),
\end{equation}
where $\ell$ is the size of the portion of the chain considered. 
For $\alpha\to1$ one obtains $S_{\mathrm{cl}}=-\ell(n_t\ln n_t+(1-n_t)\ln(1-n_t))$.  
We now observe that for the quench from $|\mathrm{F}\rangle$  the 
coherent contribution vanishes, as there are no quasiparticles. 
Thus Eq.~\eqref{renyi-yy} with $n_0=0$ (cf.~Eq~\eqref{rate-sol}) 
entirely determines the entropies (see Appendix~\ref{sec:ploss-classical}). 
This is different for the N\'eel state. We first consider the 
incoherent contribution. Within the quasiparticle 
picture this determines completely the full-system entropies, since the quasiparticles never leave the chain from its boundaries. 
Let us consider a quench from a product state obtained by translation 
of a unit cell with  $N_\uparrow$ up spins and $N_\downarrow$ 
down spins. In the absence of dissipation the spectrum of 
$G_t$ calculated over the full system is trivial at any time 
and it contains $N_\uparrow(N_\downarrow)$ eigenvalues $1(0)$. In the presence of 
dissipation the dynamics is described by the rate equation~\eqref{rate}, 
implying  that at time $t$, $G_t$ has $N_\uparrow [N_\downarrow]$ 
eigenvalues $n_t(n_0=1) [n_t(n_0=0)]$.
This implies that the full-system entropies are (see also Appendix~\ref{sec:ploss-classical})
\begin{equation}
\label{eq:ent-ave}
	\langle S_\mathrm{cl}^{(\alpha)}\rangle=
	\frac{N_\uparrow S_{\mathrm{cl}}^{(\alpha)}
	(n_0=1)+N_\downarrow 
	S_{\mathrm{cl}}^{(\alpha)}(n_0=0)}
{N_\downarrow+N_\uparrow}. 
\end{equation}
It is natural to wonder whether 
the entropy of the average $S^{(\alpha)}_\mathrm{cl}(\langle n_0\rangle)$  
plays any role. Note that since the ``melting'' of the 
initial inhomogeneity happens because of coherent hopping, 
one should expect $S^{\scriptscriptstyle (\alpha)}(\langle n_0\rangle)$ to give a 
coherent contribution.  For the following  it is useful to observe that 
$\langle S^{(\alpha)}(n_0)\rangle=S^{(\alpha)}(\langle n_0\rangle)$ 
in the limit $t\to\infty$.  

We now consider the entropies of a finite interval. 
To reveal the coherent contribution due to quasipartices it is convenient to consider 
the large $t$ limit of the von Neumann entropy. By using~\eqref{eig} 
we obtain  
\begin{multline}
\label{vn-series}
S=S_\mathrm{cl}(t\to\infty)+\sum_i\Big\{
	2b(\widetilde\lambda_i-n_\infty)\mathrm{atanh}(1-2n_\infty)\\
	+\frac{(n_\infty-\widetilde\lambda_i)^2}{2n_\infty(n_\infty-1)}
	b^2+o(b^2)\Big\},
\end{multline}
where $\widetilde \lambda_i$ are introduced in~\eqref{eig}. 
Eq.~\eqref{vn-series} can be simplified by using that  
$\sum_i\widetilde\lambda_i= \langle n_0\rangle \ell$. 
Interestingly, up to order ${\mathcal O}(b)$ Eq.~\eqref{vn-series} 
coincides with the large $t$ expansion of  $\langle S(n_0)\rangle$. 
This suggests to define the coherent part $S_\mathrm{q}$ 
as 
\begin{multline}
\label{boxed}
S_\mathrm{q}\equiv S-\langle S_\mathrm{cl}(n_0)\rangle=\\
\frac{e^{-2(\gamma^++\gamma^-)t}}{2n_\infty(n_\infty-1)}
\Big[\sum_i\Big(\frac{\widetilde\nu_i^2-1}{4}\Big) 
+\langle n_0\rangle-\langle n_0^2\rangle\Big]. 
\end{multline}
Here we have $\widetilde\nu_i\equiv2\widetilde \lambda_i-1$. 
For an initial product state, which is also eigenstate of the total number operator, one has $n_0-\langle n_0^2\rangle=0$. 
Eq.~\eqref{boxed} depends on $\widetilde\lambda_i$, 
suggesting that $S_\mathrm{q}$ is sensitive to coherent quasiparticle correlations,  
whereas the damping factor accounts for decoherence. 
We mention that  a similar cancellation of the incoherent part 
as in~\eqref{boxed} is encoded by construction in  
${\cal E}$ (see Appendix~\ref{sec:negat}), reflecting that the negativity is 
a genuine entanglement measure. 
To proceed, we observe that in 
free-fermion systems $\sum_i(\widetilde\nu_i)^p$ can be 
calculated analytically~\cite{fagotti-2012} 
for any $p$ (see Appendix~\ref{sec:mau}). For the N\'eel quench 
for $t\ll\ell$ one obtains  
\begin{equation}
\label{boxed-1}
S_\mathrm{q}\approx\frac{4t}{\pi}\frac{e^{-2(\gamma^++\gamma^-)t}}{n_\infty(1-n_\infty)}. 
\end{equation}
A similar result for $S_\mathrm{q}^{\scriptscriptstyle(\alpha)}$ is reported in  
Appendix~\ref{sec:mau}. Interestingly, Eq.~\eqref{boxed-1} 
suggests that quasiparticles have finite lifetime $1/(\gamma^++\gamma^-)$, but this is not the only effect of dissipation. Indeed, the quasi-particle correlation content is not $\ln(2)$ as in the pure case. 
Remarkably, we can resum the large $t$ expansion of Eq.~\eqref{boxed}  (see Appendix~\ref{sec:mau}). 
This yields 
\begin{multline}
\label{main-formula}
S^{(\alpha)}=\int\frac{d\lambda}{2\pi}
\Big[\mathrm{max}(1-2|v|t/\ell,0)\langle S^{(\alpha)}_\mathrm{cl}(n_0)\rangle\\
+\mathrm{min}(2|v|t/\ell,1)S^{(\alpha)}_\mathrm{cl}(\langle n_0\rangle)\Big], 
\end{multline}
with $v(\lambda)=\sin(\lambda)$ the fermion velocity. 
Eq.~\eqref{main-formula}  holds in the scaling limit 
$t,\ell\to\infty$, $\gamma^\pm\to0$ with $t/\ell$ and $\gamma^\pm\ell$,  
$\gamma^+/\gamma^-$ fixed, however, it is also quite accurate for moderately large $\gamma^\pm$ and small $\ell$. 
From~\eqref{main-formula}, the scaling behavior~\eqref{eq:scaling} 
is apparent (note that $S_\mathrm{cl}\propto\ell$). 
It is enlightening to consider the limit $v_\mathrm{max}t/\ell<1$. 
Clearly, Eq.~\eqref{main-formula} contains the 
incoherent contribution $\langle S_\mathrm{cl}(n_0)\rangle$. 
Concomitantly, the term  
$\int d\lambda/(2\pi)|v|t[S^{\scriptscriptstyle(\alpha)}_\mathrm{cl}(\langle n_0\rangle)-
\langle S^{\scriptscriptstyle(\alpha)}_\mathrm{cl}(n_0)\rangle]$, which 
coincides with $S_\mathrm{q}$ (cf.~\eqref{boxed}), describes the coherent 
contribution due to quasiparticle pairs. 
Interestingly, the second term in the square brackets reveals 
how the environment suppresses the quasi-particle correlation. 
A striking consequence of~\eqref{main-formula} is that 
$I_A^{\scriptscriptstyle(\alpha)}$ is sensitive to the coherent term only. 
If $A$ (cf.~Fig.~\ref{fig0:chain}) is the semi-infinite chain, it  
is straightforward to check that $I_A^{\scriptscriptstyle(\alpha)}=
2S^{\scriptscriptstyle(\alpha)}_\mathrm{q}$.  

Eq.~\eqref{main-formula} exhibits an interesting behavior 
at short times $t\ll 1/\gamma^{\pm}$. 
It is easy to show that  
\begin{equation}
\label{eq:short-1}
	\langle	S^{(\alpha)}_\mathrm{cl}(n_0)\rangle\to
	\left\{\begin{array}{cc}
	\frac{\alpha}{2(\alpha-1)}t\ell(\gamma^++\gamma^-) & \alpha>1\\
	\frac{t^\alpha\ell}{2(1-\alpha)}((\gamma^+)^\alpha+
	(\gamma^-)^\alpha) & \alpha<1 \\
-\frac{t\ell}{2}\sum_{a=\pm}\gamma^a(\ln t\gamma^a
	-1) & \alpha=1
\end{array}
	\right.
\end{equation}
Crucially,  $\langle S^{\scriptscriptstyle
(\alpha)}\rangle$ vanishes for $\gamma^+,\gamma^-\to 0$, revealing its 
incoherent origin. Oppositely, one has  
$S^{(\alpha)}(\langle n_0\rangle)\to \ell\ln 2$, 
as without dissipation. 
We notice that Eq.~\eqref{eq:short-1} predicts a non-linear growth 
with time for the entropy,  
even for $t\ll 1/\gamma^\pm$, in contrast with the random unitary 
scenario~\cite{li-2018,skinner-2019,jian-2019,choi-2019,cao-2019}.

\paragraph*{Numerical results.--} We now provide some numerical results. 
In Fig.~\ref{fig:quantum-entropy} (a) we focus on the von Neumann entropy. 
The dashed dotted lines denote~\eqref{main-formula}, and are in perfect 
agreement with exact numerical data. The agreement is 
excellent already for $\ell=20$, in contrast with the 
unitary case (see Ref.~\onlinecite{fagotti-2008}) where  
$t,\ell\to\infty$ is needed. 
In Fig.~\ref{fig:quantum-entropy} (b) we discuss $I_A$. 
Again, already for $\ell=20$ the numerical data are 
in spectacular agreement with theoretical predictions from Eq.~\eqref{main-formula}. 
In Fig.~\ref{fig:quantum-entropy} (c) we discuss the large $t$ limit 
of $S^{\scriptscriptstyle(\alpha)}_\mathrm{q}$ 
(cf.~\eqref{boxed}). For all the values of 
$\alpha,\gamma^\pm,\ell$, the agreement 
with the theory is perfect. 

We now discuss the negativity ${\cal E}$ 
in Fig.~\ref{fig:neg-panel}. In Fig.~\ref{fig:neg-panel} (a) we show ${\cal E}$ for 
gain/loss dissipation. The behavior is 
qualitatively similar to that of $I_A$, with a non-linear 
increase up to $t\approx 1/(\gamma^++\gamma^-)$, followed by an   
exponential decay (see the inset) at long times. This reflects that 
the Liouvillian has a gap $\gamma^++\gamma^-$. 
In Fig.~\ref{fig:neg-panel} (b) we consider the  off-diagonal gain/loss 
matrices $\Gamma^\pm_{mn}=\gamma^\pm(\gamma^*\delta_{mn}+\delta_{m,n-1}+
\delta_{m,n+1})$, where $\gamma^*\ge 2$.
Now, the Liouvillian is gapless for $\gamma^*=2$. 
Figure~\ref{fig:neg-panel} shows that for $\gamma>2$, 
${\cal E }$ decays exponentially as in (a). Surprisingly, for $\gamma^*=2$, 
${\cal E}$ attains a finite, albeit small, value at $t\to\infty$. 
For $\gamma^*>2$, the decay 
seems to be well described by ${\cal E}\propto\exp(-2(\gamma^++\gamma^-)(\gamma^*-2))t$. 

In Fig.~\ref{fig:neg-panel} (c-d) we consider incoherent hopping 
and dephasing, respectively, with rates $\gamma_h,\gamma_d$ (see Appendix~\ref{sec:inc-hop}~
\ref{sec:deph}). 
The generator is no longer quadratic, although 
$G_t$ can be obtained efficiently. 
Here we  consider $S^{\scriptscriptstyle(\alpha)}$ and ${\cal E}$ as 
defined from $G_t$, neglecting deviations from gaussian behavior.
As it is clear from Fig.~\ref{fig:neg-panel} (c-d), 
${\cal E}$  increases in a non-linear way up to $t\approx 1/\gamma$, 
and decreases for $t\to\infty$. In both cases 
${\cal E}$ attains a finite value ${\cal E}\approx 10^{-6}$ at $t\gg\ell$, which 
could be attributed to the existence of a metastable prestationary regime.
Finally, in Fig.~\ref{fig:neg-panel} (e-g) we show that 
the qualitative behavior of ${\cal E}$ and $I_A$ are similar.

A striking prediction of~\eqref{main-formula} is that $S^{\scriptscriptstyle 
(\alpha)}$ exhibits the 
scaling~\eqref{eq:scaling}. This  is inherited by $I_A^{\scriptscriptstyle(\alpha)}$. 
It is natural to expect that the quasiparticle 
picture holds for ${\cal E}$.
In Fig.~\ref{fig:negscal} we compare  
$\gamma {\cal E}$ and $\gamma I_A^{\scriptscriptstyle(1/2)}/2$ 
plotted versus $\gamma t$ ($\gamma=\gamma^\pm,\gamma_h,\gamma_d$).  
Clear scaling behavior is visible in all cases. 
Crucially, the scaling functions describing ${\cal E}$ and 
$I_A^{\scriptscriptstyle(1/2)}$ are different, unlike the unitary 
case (see Ref.~\onlinecite{alba-epl}).

\paragraph*{Conclusions.--}

We have provided the first exact formulae describing  information 
spreading in open quantum systems. There is an enourmous scope for future research. 
First, there is the need to verify our results in different models 
and for different initial states. Second, 
our results lay the foundation for generalizations of the quasiparticle picture to 
dissipative interacting 
integrable models. This would allow to study the interplay 
between interactions and dissipation~\cite{cai-2013} in the entanglement 
dynamics. Furthermore, it is 
extremely important to generalize~\eqref{main-formula} for the negativity. 
Finally, as observed, dissipation can serve to mitigate scaling corrections: this certainly deserves further investigation in numerical simulations and in experiments. 

\paragraph*{Acknowledgments.--}
We would like to thank Maurizio Fagotti for several useful 
discussions about the results of Ref.~\onlinecite{fagotti-2012}. 
V.A.~acknowledges support from 
the European Research Council under ERC Advanced grant 743032 DYNAMINT. 
F.C.~acknowledges support through a Teach@T\"ubingen Fellowship.



%
%
%
%
%
%
%
%

%
%
%
%
%
%
%
%
%
%
%
%
%
%
%
%
%
%
%
%
%
%
%
%
%
%
%
%
%
%
%
%
%
%
%
%
\begin{appendix}

\section{A simple proof of the scaling~\eqref{eq:scaling}} 
\label{sec-quasi}

In this section we show that the scaling behavior~\eqref{eq:scaling} 
\begin{equation}
\gamma {\cal E}=g(\gamma t,\gamma\ell) 
\end{equation}
can be derived by assuming that the quasiparticle picture holds and 
that the quasiparticles have a finite lifetime. 

We focus on the negativity ${\cal E}$ between subsystem $A$ and the 
rest (see Fig.~\ref{fig0:chain}), considering the limit $L\to\infty$. 
Within the quasiparticle picture, the negativity is proportional to 
the number of quasiparticles that are shared between $A$ and its 
complement. This implies that 
\begin{multline}
	{\cal E}=\int_{0}^\pi d\lambda\int_0^L dx\int_{x'\in A}  dx'
	\int_{x''\in \bar A} dx'' \Big
	\{\\e(\lambda)\delta(x'-x-v(-\lambda)t)\delta(x''-x-v(\lambda)t)
	\Big\},
\end{multline}
where $x$ is the point from which the entangled pair is emitted, and 
$\lambda$ is the quasimomentum. 
Note that we perform the integral over half of the Brillouin 
zone $\lambda\in[0,\pi]$. 
We definite the two intervals as $A=[0,\ell]$ and $\bar A=[\ell+1,L]$. Here 
$e(\lambda)$ is the contribution of the quasiparticles to the negativity. 
We can perform the integration over $x'$ to obtain 
\begin{multline}
{\cal E}=\int_{0}^\pi d\lambda\int_0^L dx
	\int_{x''\in \bar A} dx'' \Big\{\\
	e(\lambda)\theta(x+v(-\lambda)t+\ell)
	\theta(\ell-x-v(-\lambda)t)\delta(x''-x-v(\lambda)t)\Big\}. 
\end{multline}
Here the term $\theta(x+v(-\lambda)+\ell)$ takes into account that the 
quasiparticles can bounce on the origin and exit subsystem $A$ at 
later times. 
The integration over $x''$ gives 
\begin{multline}
{\cal E}=\int_{0}^\pi d\lambda\int_0^L dx
	\Big\{e(\lambda)\theta(x-v(\lambda)t+\ell)
	\theta(\ell-x+v(\lambda)t)\\\times\theta(x+v(\lambda)t-\ell)\theta(L-x-v(\lambda)t)
\Big\},
\end{multline}
where we used that $v(\lambda)=-v(\lambda)$. 
We now assume that $L\to\infty$. This implies that $\theta(L-x-v(\lambda)t)=1$ 
for any $\lambda, t$. 
Thus, the expression above becomes
\begin{equation}
\label{quasi-e}
{\cal E}=2\int_0^\pi d\lambda \min(v(\lambda)t,\ell)e(\lambda)
\end{equation}
In the presence of the environment, we assume that 
the probability for a quasiparticle with quasimomentum $\lambda$ to survive 
up to time $t$ is $\exp(-\gamma t)$, with $\gamma$ the 
relevant rate. The calculations leading to~\eqref{quasi-e} 
remain the same, the only modification is a multiplicative factor 
$\exp(-2\gamma t)$, which is the probability for both the members of the 
entangled pair to survive. Clearly, now~\eqref{quasi-e} satisfies the 
scaling~\eqref{eq:scaling}. 

%
%

\section{Some numerical checks for gain/loss Lindbladians}
\label{sec:ploss-classical}

\begin{figure}[t]
\includegraphics[width=0.4\textwidth]{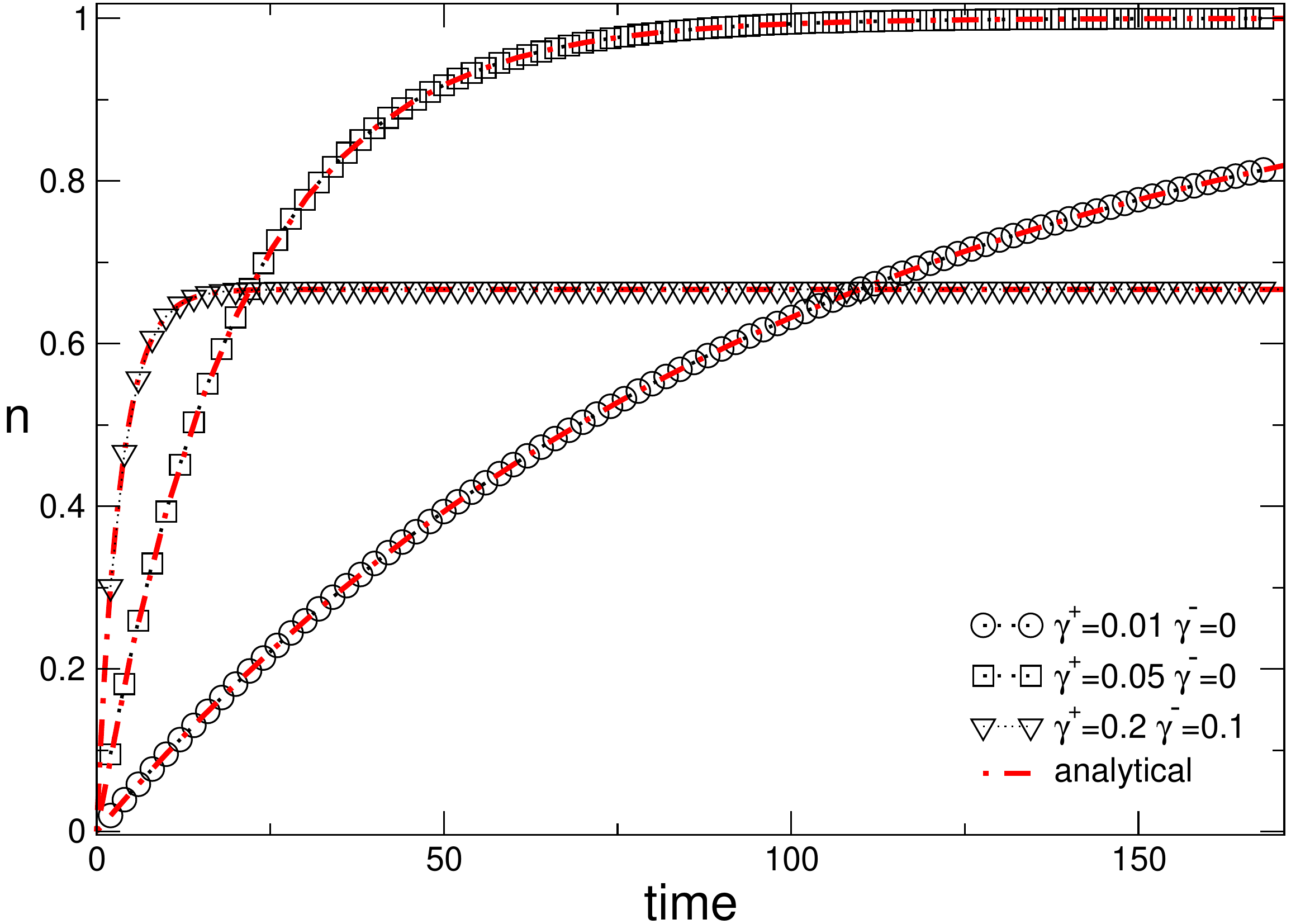}
\includegraphics[width=0.403\textwidth]{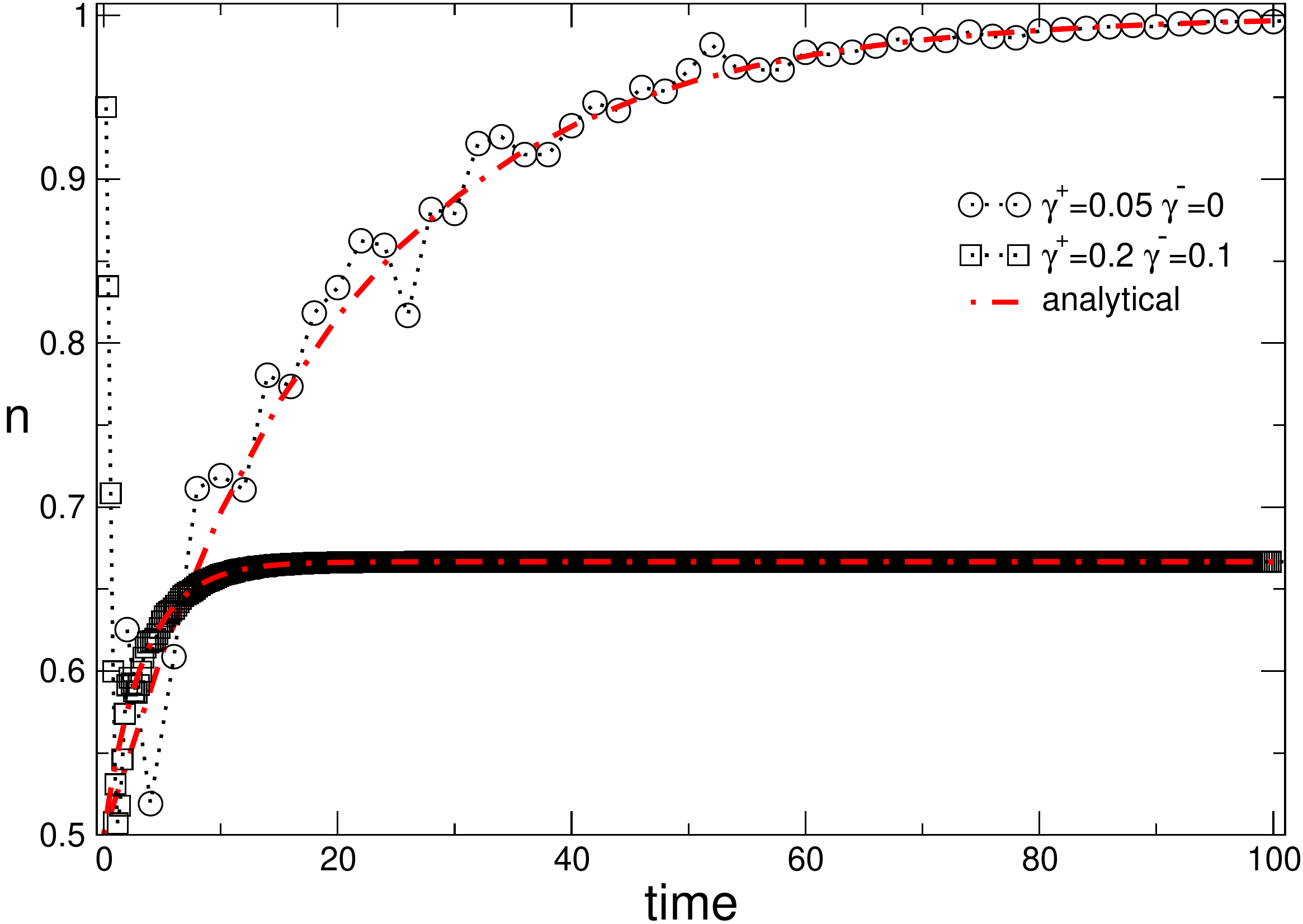}
\caption{ Dynamics of the local density of fermions $n$ in the 
 XX chain in the presence 
 of linear diagonal dissipation. (top) Results for the 
 dynamics starting from the ferromagnetic state. (bottom) The same as 
 in (top) for the quench from  the N\'eel state. In both panels the 
 symbols are exact numerical results,  the dashed-dotted lines 
 are the analytical results in~\eqref{rate-sol}. 
 Note also that for $\gamma^+=0.05$ and $\gamma^-=0$ 
 strong oscillations are present for the N\'eel quench. 
}
\label{fig:density}
\end{figure}
%
In this section we provide some numerical checks of~\eqref{rate-sol}. We also 
discuss the theory prediction for the full-system von Neumann entropy $S_T$, 
which depends only on the incoherent contribution. 

In Fig.~\ref{fig:density} we compare Eq.~\eqref{rate-sol} with exact numerical data 
obtained by using~\eqref{CovMat}. In the top panel we consider the quench 
from the ferromagnetic state $|\mathrm{F}\rangle$ in the XX chain. 
We also consider the quench from the N\'eel state (bottom panel). 
For $|\mathrm{F}\rangle$ 
the initial correlation matrix is exactly zero, whereas for the 
N\'eel state it is given as $G_0=(1-(-1)^i)/2\delta_{ij})$. 
In both panels, the numerical data obtained from~\eqref{CovMat} are 
well described by~\eqref{rate-sol}. We note that for the quench from the 
N\'eel state and $\gamma^+=0.05$ there are oscillating corrections 
to~\eqref{rate-sol}, which disappear in the long-time limit. 
These signals that for small $\gamma^\pm$ the system remain coherent for 
short times.

\begin{figure}[t]
\includegraphics[width=0.4\textwidth]{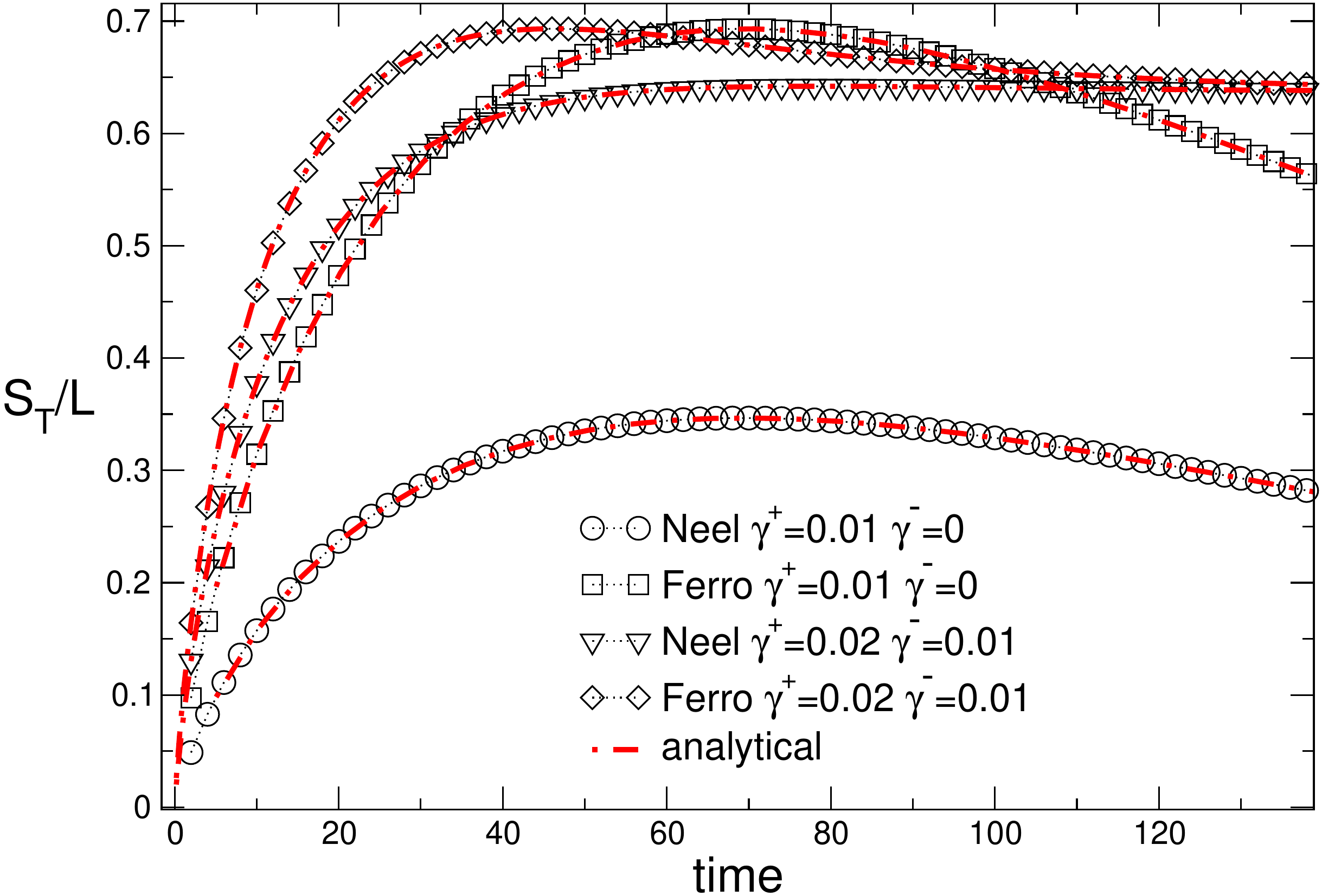}
\caption{ Dynamics of the entanglement entropy of the full system in the presence of 
 gain/loss. The system is the XX chain. The data are for the 
 quench from the N\'eel state (circles and triangles) and the 
 ferromagnetic state (diamonds and squares). The dashed dotted lines denote 
 the analytical results obtained from~\eqref{renyi-yy} and~\eqref{rate-sol}. 
}
\label{fig:entT}
\end{figure}

The behavior of the full-system entropy is checked in Fig.~\ref{fig:entT}. 
The figure shows the density of 
entropy $S_T/L$ of the full chain for the quenches from $|\mathrm{F}\rangle$ 
and the N\'eel state. For $|\mathrm{F}\rangle$ the von Neumann entropy is 
obtained by using~\eqref{rate-sol}, with  $n_0=0$ in~\eqref{renyi-yy}, whereas 
for the N\'eel state we used~\eqref{eq:ent-ave}. 

\section{Two-point functions for incoherent hopping}
\label{sec:inc-hop}

In this section we review the approach to determine the two-point 
correlation function $G=\langle c_n^\dagger c_m\rangle$ in free-fermion 
systems. We follow Ref.~\onlinecite{eisler-2011}. 

Let us consider Majorana fermions $a_{m}$ as 
\begin{align}
& a_{2m-1}=c_m+c^\dagger_m\\
& a_{2m}=i(c_m-c^\dagger_m)
\end{align}
with anticommutation relations $\{a_k,a_l\}=2\delta_{kl}$. 
We consider the generic Hamiltonian 
\begin{equation}
	H=\frac{i}{4}\sum_{k,l}H_{kl}a_ka_l,\quad H_{kl}=-H_{lk}
\end{equation}
We also consider the generic Lindblad operator $L_\alpha$ 
\begin{equation}
\label{eq:lin-inc}
	L_\alpha=\frac{i}{4}\sum_{k,l}L_{\alpha,kl}a_ka_l
\end{equation}
It is important for the following to define a generic 
{\it ordered} string of Majorana operators $\Gamma_{\underline{\nu}}$ as 
\begin{equation}
\Gamma_{\underline{\nu}}=a_1^{\nu_1}a_2^{\nu_2}\dots a_{2L}^{\nu_{2L}}
\end{equation}
Here $\nu_i\in\{0,1\}$ are the occupation numbers in the string. 
We also define Majorana {\it superoperators} $\hat a_{j}$ with action 
\begin{align}
& \hat a_j\Gamma_{\underline{\nu}}=\delta_{1,\nu_j}\pi_j\Gamma_{\underline{\nu}'}\\
& \hat a_j^\dagger\Gamma_{\underline{\nu}}=\delta_{0,\nu_j}\pi_j\Gamma_{\underline{\nu}'},
\end{align}
where $\nu_j'=1-\nu_j$. The phase factor 
\begin{equation}
\pi_j=\exp\Big(i\pi\sum_{k=1}^{j-1}\nu_k\Big)
\end{equation}
ensures the fermionic anticommutation relations $\{\hat a_k,\hat a_l^\dagger\}=
\delta_{kl}$. It is easy to prove that 
\begin{equation}
\label{comm-kin}
[a_ka_l,\Gamma_{\underline{\nu}}]=2(\hat a_k^\dagger \hat 
a_l-\hat a_l^\dagger\hat a_k)\Gamma_{\underline{\nu}}
\end{equation}
For instance, the r.h.s. of~\eqref{comm-kin} becomes 
\begin{equation}
2(-1)^{l-k}\pi_k\pi_l(\delta_{1,\nu_l}\delta_{0,\nu_k}\Gamma_{\underline{\nu}'}
+\delta_{0,\nu_l}\delta_{1,\nu_k}\Gamma_{\underline{\nu}''})
\end{equation}
where $\nu_l'=0=1-\nu'_k$ and $\nu_l''=1=1-\nu_k''$. 
It is easy to check that if $\nu_l=\nu_k$ the term $a_k a_l$ 
commutes with $\Gamma_{\underline{\nu}}$. If $\nu_l\ne\nu_k$ one has 
that 
\begin{equation}
a_ka_l\Gamma_{\underline{\nu}}=-\Gamma_{\underline{\nu}}a_ka_l, 
\end{equation}
which allows to prove~\eqref{comm-kin}. By using~\eqref{comm-kin} 
one can derive the coherent part of the evolution as 
\begin{equation}
i[H,\Gamma_{\underline{\nu}}]=-\sum_{k,l}H_{kl}\hat a_k^\dagger \hat a_l.
\end{equation}
The incoherent contributions in~\eqref{l-eq} can be calculated as 
well. By using~\eqref{eq:lin-inc} one obtains 
\begin{multline}
	{\mathcal L}_\mathrm{d}(\Gamma_{\underline{\nu}})\equiv
\frac{1}{16}\sum_\alpha\Big(L_\alpha^\dagger
	\Gamma_{\underline{\nu}}L_\alpha-
	\frac{1}{2}\{L^\dagger_\alpha L_\alpha,\Gamma_{\underline{\nu}}\}\Big)=\\
	-\frac{1}{32}\sum_{\alpha,ijkl}\Big(2L_{\alpha,ij}a_ia_j\Gamma_{\underline{\nu}}
	L_{\alpha,kl}a_ka_l\\-
	\{L_{\alpha,ij}L_{\alpha,kl}a_ia_ja_ka_l,\Gamma_{\underline{\nu}}\}\Big),
\end{multline}
where we used that $L_{\alpha,ij}=-L_{\alpha,ji}$. 
Simple manipulations yield 
\begin{equation}
\label{eq:idd}
	{\mathcal L}_\mathrm{d}(\Gamma_{\underline{\nu}})=\frac{1}{32}\sum_{\alpha,ijkl}
	L_{\alpha,ij}L_{\alpha,kl}[a_ia_j,[a_ka_l,\Gamma_{\underline{\nu}}]]. 
\end{equation}
By applying~\eqref{comm-kin}, one has 
\begin{equation}
	{\mathcal L}_\mathrm{d}(\Gamma_{\underline{\nu}})=
	\frac{1}{2}\sum_{\alpha,ijkl}
	L_{\alpha,ij}L_{\alpha,kl}\hat a^\dagger_i \hat a_j\hat a_k^\dagger \hat a_l 
	\Gamma_{\underline{\nu}}. 
\end{equation}
By using that $L_{ij}=-L_{ji}=-L^T_{ij}$, and upon normal ordering, one obtains 
that the  Liouvillian in~\eqref{l-eq} is given as 
\begin{equation}
	\label{eq:vikt-id}
	{\mathcal L}=-\sum_{kl}\widetilde H_{kl}\hat a^\dagger_k\hat a_l
	+\frac{1}{2}\sum_\alpha\sum_{ijkl}L^T_{\alpha,ij}L_{\alpha,kl}\hat a^\dagger_i 	\hat a_k^\dagger \hat a_j\hat a_l, 
\end{equation}
where we defined $\widetilde H_{kl}$ as 
\begin{equation}
\widetilde H_{kl}=H_{kl}+\frac{1}{2}\sum_\alpha L^{T}_{\alpha,kl}L_{\alpha,kl}. 
\end{equation}

Let us now consider the $XX$ hamiltonian. In terms of Majorana fermions one has 
\begin{equation}
	\label{eq:xx-ham-1}
	H_{kl}=-(\delta_{k,l-1}+\delta_{k,l+1})\otimes
	\left(\begin{array}{cc}0 & -1\\
		   	       1 & 0\end{array}\right)
\end{equation}
We choose the Lindbladians corresponding to incoherent hopping~\cite{eisler-2011} 
\begin{multline}
	L_{2j-1,kl}=\sqrt{\frac{\gamma_\mathrm{h}}{2}}(\delta_{k,j}\delta_{l,j+1}\\
\qquad+\delta_{k,j+1}\delta_{l,j})
\otimes\left(\begin{array}{cc}
0 & -1\\
1 & 0
\end{array}\right)
\end{multline}
\begin{align}
	& L_{2j,kl}=\sqrt{\frac{\gamma_\mathrm{h}}{2}}(\delta_{k,j}\delta_{l,j+1}-\delta_{k,j+1}\delta_{l,j})
\otimes\left(\begin{array}{cc}
1 & 0\\
0 & 1
\end{array}\right)\\
& \frac{1}{2}\sum_\alpha L_\alpha^T L_\alpha=\delta_{kl}
\otimes\left(\begin{array}{cc}
1 & 0\\
0 & 1
\end{array}\right)
\end{align}
It is convenient to work with the fermionic superoperators 
\begin{align}
\label{eq:tr-1}
	& 	\hat{{a}}_{-,m}=\frac{1}{\sqrt{2}}(\hat a_{2m-1}-i\hat a_{2m}),\\
\label{eq:tr-2}
	& 	\hat a_{+,m}=\frac{1}{\sqrt{2}}(\hat a_{2m-1}+i\hat a_{2m})
\end{align}
which diagonalize the two-by-two matrix appearing in~\eqref{eq:xx-ham-1} 
with eigenvalues $\pm i$. The inverse of~\eqref{eq:tr-1}\eqref{eq:tr-2} is 
\begin{align}
	& \hat a_{2j-1}=\frac{1}{\sqrt{2}}(\hat a_{-,m}+\hat a_{+,m}),\\
	& \hat a_{2j}=\frac{i}{\sqrt{2}}(\hat a_{-,m}-\hat a_{+,m}). 
\end{align}
Now, from~\eqref{eq:vikt-id} one obtains 
\begin{equation}
	{\mathcal L}={\mathcal L}_\mathrm{+}+{\mathcal L}_\mathrm{-}+
	{\mathcal L}_\mathrm{+-}. 
\end{equation}
Here we defined 
\begin{multline}
{\mathcal L}_\pm=
\sum_{m=1}^L\Big[i(\hat a^\dagger_{\pm,m}\hat a_{\pm,m+1}+\hat a_{\pm,m+1}^\dagger 
	\hat a_{\pm,m})-\gamma_\mathrm{h}\hat a^\dagger_{\pm,m}\hat a_{\pm,m}\\+
	\gamma_\mathrm{h} \hat a_{\pm,m}^\dagger\hat a_{\pm,m}\hat a_{\pm,m+1}^\dagger 
	\hat a_{\pm,m+1}
\Big]
\end{multline}
\begin{multline}
	{\mathcal L}_{+-}=\sum_{m=1}^L\gamma_\mathrm{h}\Big[\hat a_{-,m}^\dagger \hat a_{-,m+1}\hat a_{+,m}^\dagger
	\hat a_{+,m+1}\\+\hat a^\dagger_{+,m+1}\hat a_{+,m}\hat 
a^\dagger_{-,m+1}\hat a_{-,m}\Big]
\end{multline}
To proceed, one has to determine the 
action of the fermionic superoperators 
on a string of fermions. 
A simple calculation gives that the only nonzero combinations 
are 
\begin{align}
	& \hat a_{-,k} a^\dagger_m=\frac{1}{\sqrt{2}}\delta_{km}\\
	& \hat a_{+,k} a_m=\frac{1}{\sqrt{2}}\delta_{k,m}\\
	& \hat a_{-,k}^\dagger=\sqrt{2}a^\dagger_k\\
	& \hat a_{+,k}^\dagger=\sqrt{2}a_k.
\end{align}
Now one can derive the evolution of $G_{nm}=\langle a^\dagger_n a_m\rangle$ by using 
that 
\begin{equation}
	\frac{d}{dt}G_{kl}=\langle{\mathcal L}(a^\dagger_k a_l)\rangle.
\end{equation}
One obtains the system of equations as 
\begin{multline}
\label{eq:inc-hop-ev}
\frac{d}{dt}G_{kl}=i(G_{k-1,l}+G_{k+1,l}-G_{k,l-1}-G_{k,l+1})\\-2\gamma_\mathrm{h} G_{k,l}
	+\gamma_\mathrm{h}\delta_{k,l}(G_{k-1,k-1}+G_{k+1,k+1}). 
\end{multline}
%

\begin{figure}[t]
\includegraphics[width=0.4\textwidth]{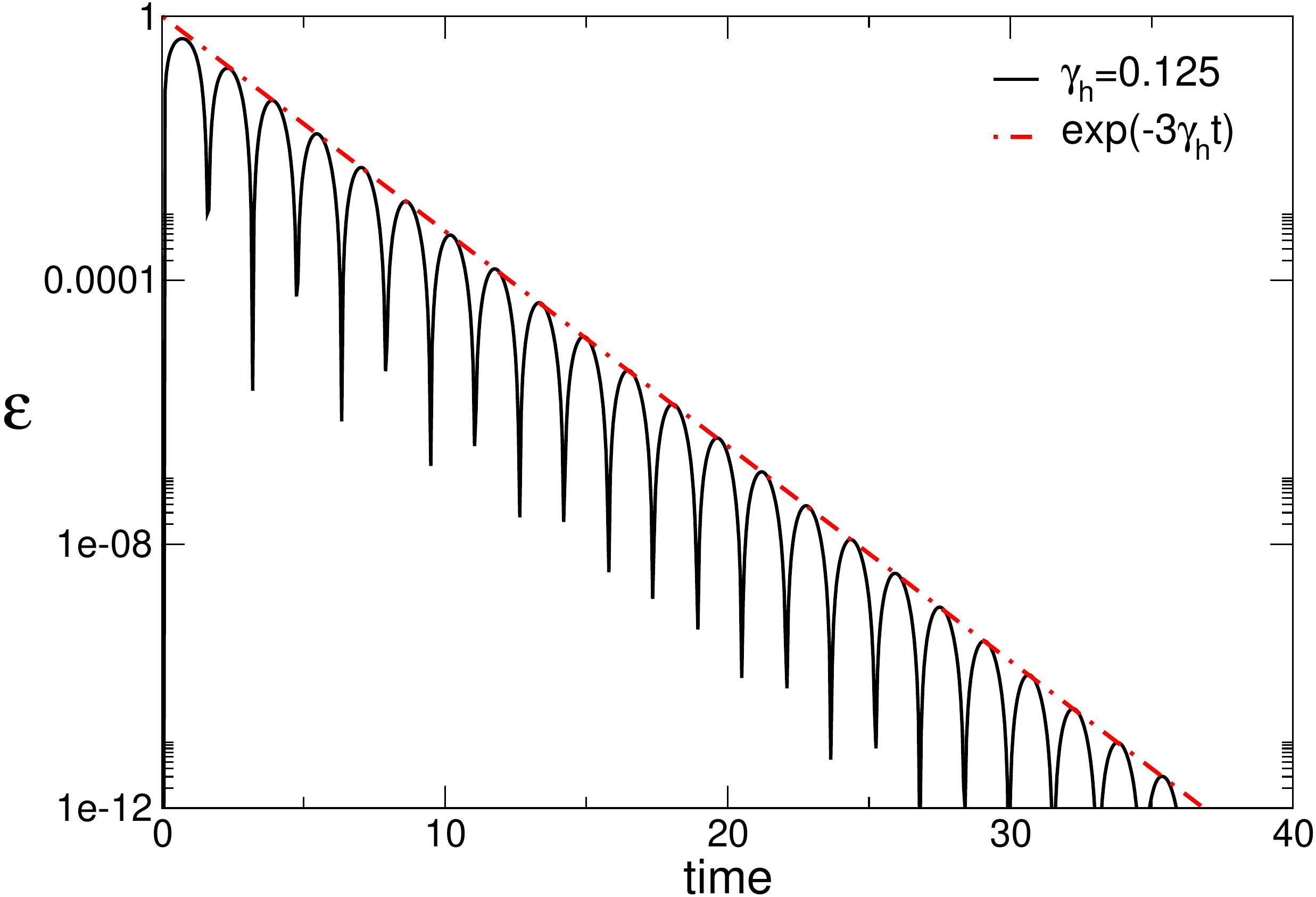}
\caption{ 
 Dynamics of the negativity between two spins in the $XX$ chain 
 with incoherent hopping with rate $\gamma_h$. The initial 
 state is the N\'eel state. The continuous 
 line denotes exact results. The dashed-dotted line is $\propto 
 e^{-3\gamma_\mathrm{h} t}$. 
}
\label{fig:two-hop}
\end{figure}

\subsection{The case of two qbits} 

It is interesting to study the logarithmic negativity of a two-qbits 
system. Here we consider the vectorized correlation matrix $G_k$. 
From~\eqref{eq:inc-hop-ev}, the evolution of $G_k$ is determined by 
\begin{equation}
\label{eq:gt-1}
	\frac{d G_{k}}{dt}=\sum_l M_{kl} G_l, 
\end{equation}
with the matrix $M_{kl}$ given as 
\begin{equation}
	M_{kl}=\left(\begin{array}{cccc}
		-2\gamma_\mathrm{h} & -i & i & 2\gamma_\mathrm{h}\\
		-i & -2\gamma_\mathrm{h} & 0 & i\\
		i & 0 & -2\gamma_\mathrm{h} & -i \\
		2\gamma_\mathrm{h} & i & -i & -2\gamma_\mathrm{h}
	\end{array}\right)
\end{equation}
Note that $M$ is complex and symmetric. 
The eigenvalues of $M$ are given as 
\begin{align}
&\lambda_0=0\\
\label{eig-1}
&\lambda_1=-2\gamma_\mathrm{h}\\
&\lambda_\pm=-3\gamma_\mathrm{h}\pm\sqrt{\gamma_\mathrm{h}^2-4}
\end{align}
with eigenvectors (not normalized)
\begin{align}
& v_0=(1,0,0,1)\\
& v_1=(0,1,1,0)
\end{align}
\begin{multline}
	v_-=(-1,i(-\gamma_\mathrm{h}+\sqrt{\gamma_\mathrm{h}^2-4})/2,\\
	-i(-\gamma_\mathrm{h}+\sqrt{\gamma_\mathrm{h}^2-4})/2,1)
\end{multline}
\begin{multline}
	v_+=(-1,i(-\gamma_\mathrm{h}-\sqrt{\gamma_\mathrm{h}^2-4})/2,\\
	-i(-\gamma_\mathrm{h}-\sqrt{\gamma_\mathrm{h}^2-4})/2,1)
\end{multline}
The matrix $M$ is diagonlized as $D=UMU^T$ by the matrix 
$U_{ij}= \tilde v_{ij}$. The eigenvectors are normalized 
as $\tilde v_i=v_i/(\sum_{j}v_{ij}^2)^{1/2}$. 
The solution of~\eqref{eq:gt} is obtained 
\begin{equation}
\label{eq:rico}
G(t)=\sum_l M_{kl}(t)G_l(0), 
\end{equation}
where 
\begin{multline}
\label{eq:mtt}
	M(t)=e^{\lambda_1 t}|\tilde v_1\rangle\langle \tilde v_1^*|+
	e^{\lambda_+ t}|\tilde v_+\rangle\langle \tilde v_+^*|\\+
	e^{\lambda_- t}|\tilde v_-\rangle\langle \tilde v_-^*|. 
\end{multline}
To proceed one has to apply $M(t)$ to the vector with the initial 
correlations. From the time-evolved correlators it is straightforward 
to obtain the negativity associated with the bipartition in which 
$A$ is one of the two spins (see Appendix~\ref{sec:negat}). 

There are several interesting observations. First, there is a 
regime for  $\gamma_\mathrm{h}<2$ (weak incoherence) where quantum coherence 
to a certain extent survives. This is reflected in an oscillating 
behavior of ${\cal E}$ with period $\pi/(4-\gamma_\mathrm{h}^2)^{1/2}$, 
superposed to an exponential decay. On the other hand, for 
$\gamma_\mathrm{h}>2$ (strong incoherence), the decay of ${\cal E}$ is 
purely exponential. Finally, we should observe that the exponential 
decay is ${\cal E}\propto e^{-3\gamma_\mathrm{h}t}$, and not 
${\cal E}\propto e^{-2\gamma_\mathrm{h}t}$, as one would have expected 
from~\eqref{eig-1}. This is because of the structure of the N\'eel state. 
This features are illustrated in Fig.~\ref{fig:two-hop}.

\section{Two-point function for dephasing}
\label{sec:deph}

We now consider dephasing noise. 
Now the Lindblad operators are given as 
\begin{equation}
	L_{Lj}=L_{Rj}=\sqrt{\gamma_\mathrm{d}} c_j^\dagger c_j
\end{equation}
Note that the Lindbladian is hermitian, in contrast with the case of 
incoherent hopping. In terms of Majorana fermions one can write
\begin{multline}
	\sqrt{\gamma_\mathrm{d}} c_j^\dagger c_j=\frac{\sqrt{\gamma_\mathrm{d}}}
	{4}(a_{2j}a_{2j}
	+a_{2j-1}a_{2j-1}\\-i a_{2j-1}a_{2j}+
	i a_{2j}a_{2j-1}). 
\end{multline}
Therefore we can write the Lindblad operator 
\begin{equation}
	{L}_{\alpha}=\frac{i}{4}\sum_{k,l}L_{\alpha,kl}a_ka_l, 
\end{equation}
with 
\begin{multline}
	L_{2\alpha-1,kl}=\sqrt{\gamma_\mathrm{d}}(-i\delta_{k,2\alpha-1}\delta_{l,2\alpha-1}
	\\-i\delta_{k,2\alpha}\delta_{l,2\alpha}-\delta_{k,2\alpha-1}\delta_{l,2\alpha}+
	\delta_{k,2\alpha}\delta_{l,2\alpha-1})\\
\end{multline}
\begin{equation}
	L_{2\alpha, kl}=0.
\end{equation}
Eq.~\eqref{eq:idd} holds for the dephasing as well. 
By using~\eqref{comm-kin}, it is straightforward to obtain that 
\begin{multline}
	{\mathcal L}_\mathrm{d}(\Gamma_{\underline{\nu}})=\frac{1}{8}
	\sum_{ijkl}L_{\alpha,ij}L_{\alpha,kl} (
	\hat a_i^\dagger\hat a_j\hat a_k^\dagger \hat a_l
	-\hat a_i^\dagger\hat c_j\hat a_l^\dagger \hat a_k
	\\
	-\hat a_j^\dagger\hat a_i\hat a_k^\dagger \hat a_l
	+\hat a_j^\dagger\hat a_i\hat a_l^\dagger \hat a_k
	)\Gamma_{\underline{\nu}}. 
\end{multline}
One can use that $L_{\alpha,lk}=-L_{\alpha,kl}^*$. This allows one to write
\begin{multline}
\label{eq:rem}
	{\mathcal L}_\mathrm{d}(\Gamma_{\underline{\nu}})=\frac{1}{8}
	\sum_{ijkl}(L_{\alpha,ij}L_{\alpha,kl}+L_{\alpha,ij}L^*_{\alpha,kl}
	\\+L_{\alpha,ij}^*L_{\alpha,kl}+L^*_{\alpha,ij}L^*_{\alpha,kl}) 
	\hat a_i^\dagger\hat a_j\hat a_k^\dagger \hat a_l
\end{multline}
It is convenient to use the 
operators $\hat a_{\pm,m}$ defined 
in~\eqref{eq:tr-1} and~\eqref{eq:tr-2}
Thus, Eq.~\eqref{eq:rem} becomes  
\begin{multline}
{\mathcal L}_\mathrm{d}(
\Gamma_{\underline{\nu}})=\frac{\gamma_\mathrm{d}}{2}\sum_{m=1}^L
(
-\hat a^\dagger_{-,m} \hat a_{-,m}\hat a^\dagger_{-,m}\hat a_{-,m}\\
-\hat a^\dagger_{+,m} \hat a_{+,m}\hat a^\dagger_{+,m}\hat a_{+,m}
+\hat a^\dagger_{+,m} \hat a_{+,m}\hat a^\dagger_{-,m}\hat a_{-,m}\\
+\hat a^\dagger_{-,m} \hat a_{-,m}\hat a^\dagger_{+,m}\hat a_{-,m}
)\Gamma_{\underline{\nu}}
\end{multline}
The expression above can be rewritten as 
\begin{multline}
{\mathcal L}_\mathrm{d}(\Gamma_{\underline{\nu}})=
\frac{\gamma_\mathrm{d}}{2}\sum_{m=1}^L
(
-\hat a^\dagger_{-,m}\hat a_{-,m}
-\hat a^\dagger_{+,m}\hat a_{+,m}
\\
+2\hat a^\dagger_{+,m} \hat a_{+,m}\hat a^\dagger_{-,m}\hat a_{-,m}
)\Gamma_{\underline{\nu}}
\end{multline}
Finally, the evolution of the two-point correlation 
in the presence of dephasing noise is given as 
\begin{multline}
\label{eq:dephasing}
	\frac{d}{dt}G_{kl}=i(G_{k-1,l}+G_{k+1,l}-G_{k,l-1}-G_{k,l+1})\\
	-\gamma_\mathrm{d} (G_{kl}-\delta_{kl}G_{kk}).
\end{multline}
As expected, the effect of the dephasing is to suppress off-diagonal 
correlations. 

\subsection{The case of two qbits}

As for the case of incoherent hopping it is enlightening to consider the 
negativity in a two-qbits system. The evolution of the vectorized 
correlation matrix $G_k$ is
\begin{equation}
	\frac{d}{dt}G_k=\sum_l M_{kl}G_l. 
\end{equation}
Now the matrix $M_{kl}$ reads as 
\begin{equation}
	M_{kl}=\left(\begin{array}{cccc}
	0 & -i & i & 0\\
	-i & -\gamma & 0 & i\\
	i & 0 & -\gamma & -i \\
	0 & i & -i & 0
	\end{array}\right)
\end{equation}
The eigenvalues of $M$ are given as 
\begin{align}
&\lambda_0=0\\
&\lambda_1=-\gamma\\
&\lambda_\pm=(-\gamma\pm\sqrt{\gamma^2-16})/2
\end{align}
with (not normalized) eigenvectors 
\begin{align}
& v_0=(1,0,0,1)\\
& v_1=(0,1,1,0)
\end{align}
\begin{multline}
v_-=(-1,i(\gamma+\sqrt{\gamma^2-16})/4,\\
-i(\gamma+\sqrt{\gamma^2-16})/4,1)
\end{multline}
\begin{multline}
 v_+=(-1,i(\gamma-\sqrt{\gamma^2-16})/4,\\
-i(\gamma-\sqrt{\gamma^2-16})/4,1)
\end{multline}
As for the case of incoherent hopping (see~\eqref{eq:rico}) 
the time-evolved correlation matrix is obtained from 
the  matrix $M(t)$ given as 
\begin{multline}
	M(t)=e^{\lambda_1 t}|\tilde v_1\rangle\langle \tilde v_1^*|+
	e^{\lambda_+ t}|\tilde v_+\rangle\langle \tilde v_+^*|\\
	+e^{\lambda_- t}|\tilde v_-\rangle\langle \tilde v_-^*|, 
\end{multline}
where $\tilde v_i$ are the normalized eigenvectors, as in~\eqref{eq:mtt}. 
%
\begin{figure}[t]
\includegraphics[width=0.4\textwidth]{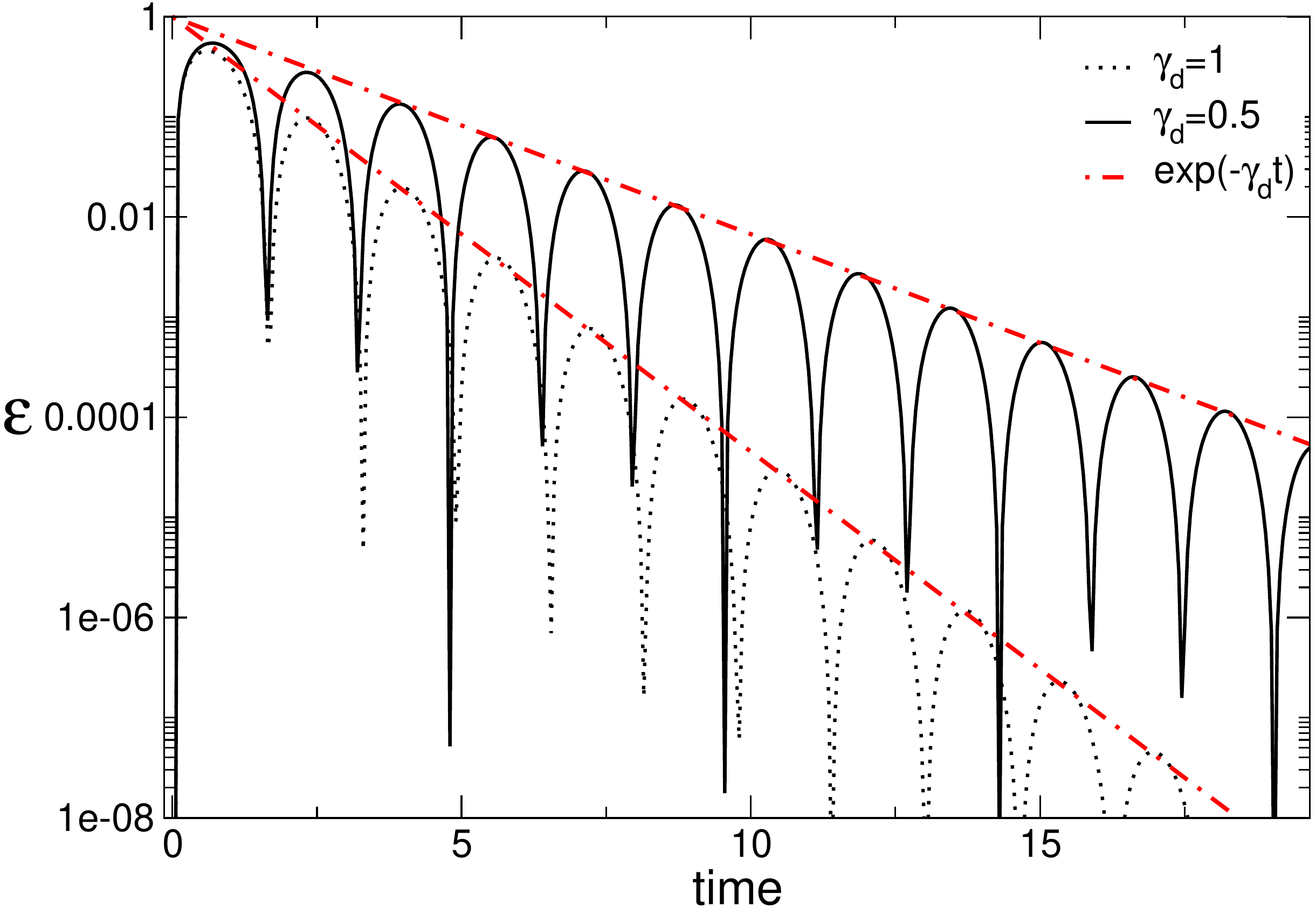}
\caption{ 
 Dynamics of the negativity ${\cal E}$ between two spins in the 
 $XX$ chain with two sites in the presence of dephasing dissipation 
 with rate $\gamma_d$. The initial state is the N\'eel state. 
 The continuous lines are exact results. The dashed-dotted line is 
 the behavior $e^{-\gamma_\mathrm{d} t}$. 
}
\label{fig:two-deph}
\end{figure}
%
Some qualitative features of the negativity ${\cal E}$ is illustrated 
in Fig.~\ref{fig:two-deph}. Now the negativity decays as 
${\cal E}\propto e^{-\gamma_\mathrm{d}t}$. Moreover, as for the incoherent 
hopping there are oscillations for $\gamma_\mathrm{d}<4$, whereas only 
exponential decay for $\gamma_\mathrm{d}\ge 4$.

\section{Additional data for incoherent hopping}

Here we present some additional data for the behavior of the negativity ${\cal E}$ 
in the $XX$ chain in the presence of incoherent hopping. Specifically, 
here we show that in the long time regime $t\gg\ell$ with $\ell$ the size 
of subsystem $A$, ${\cal E}$ attains a finite although small value. 

\begin{figure}[t]
\includegraphics[width=0.4\textwidth]{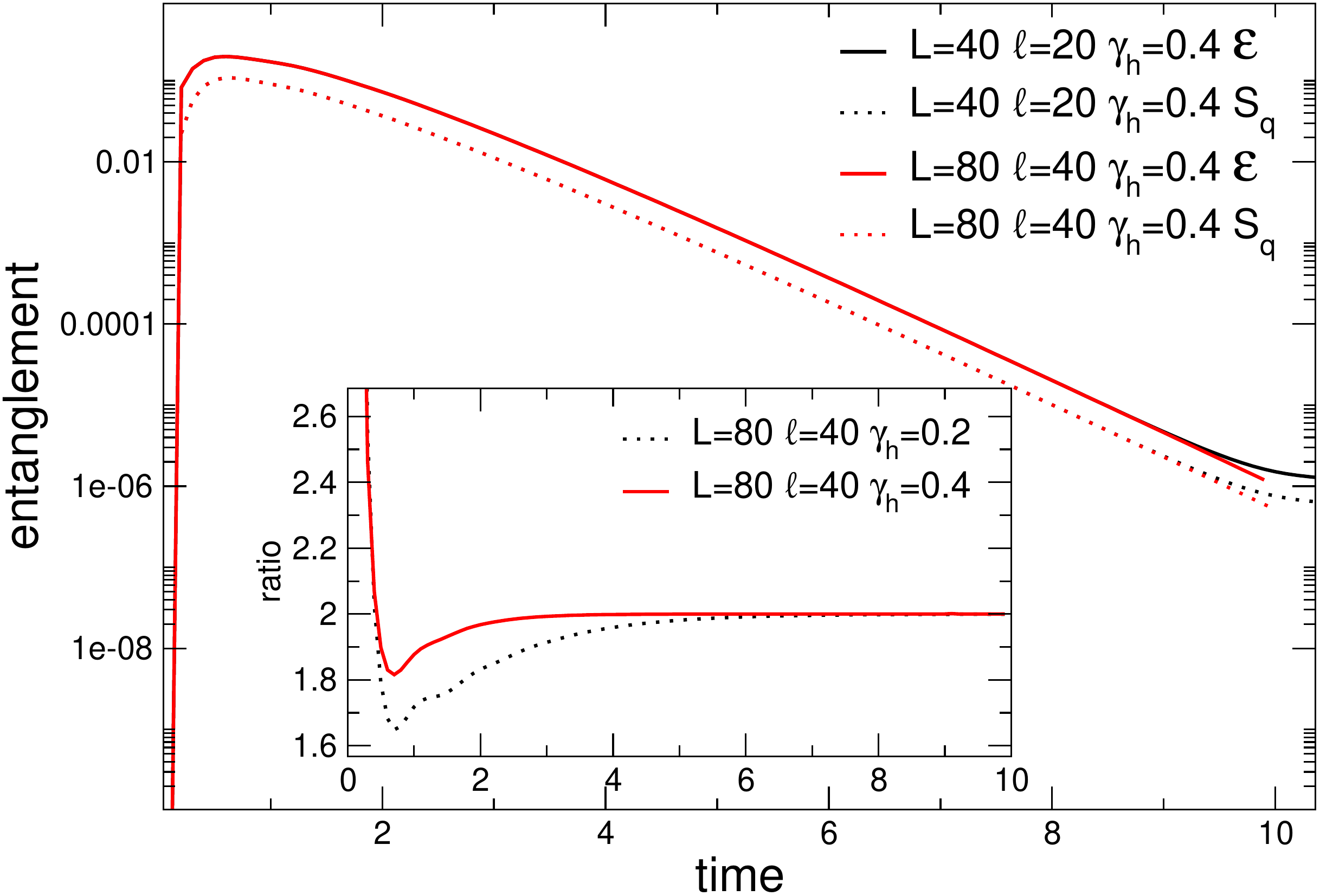}
\caption{ Dynamics of the negativity ${\cal E}$ after the quench from the 
 N\'eel state  in the $XX$ in the presence of incoherent hopping with 
 rate $\gamma_\mathrm{h}$. The continuous lines denote the negativity, 
 whereas the dotted lines are for $S_\mathrm{q}$ (cf.~\eqref{boxed}). 
 The inset show the ratio between the two. Note that at long times 
 for $t\gg\ell$ both ${\cal E}$ and $S_\mathrm{q}$ attain a 
 finite small value $10^{-6}$. 
}
\label{fig:hop}
\end{figure}

Our data for several sizes $\ell$ of subsystem $A$  and of the chain 
$L$ are reported in Fig.~\ref{fig:hop}. The figure shows both the 
negativity ${\cal E}$ and the coherent contribution of the von 
Neumann entropy $S_\mathrm{q}$ (cf.~\eqref{boxed}). The latter is 
obtained by subtracting from $S$ the incoherent contribution 
(cf.~\eqref{eq:ent-ave}). Both ${\cal E}$ and $S_\mathrm{q}$ 
exhibit a similar decay. As it is clear from the inset 
one has that ${\cal E}/S_\mathrm{q}\approx 2$ for large times.  
We observe that both quantities attain a finite 
value at long times $t\gg\ell$. For instance, the data for $\ell=20$ 
seem to saturate at $t\approx 10$ to a value $10^-6$. Interestingly, 
the saturation happens later for larger values of $\ell=40$.

\section{Fermionic logarithmic negativity}
\label{sec:negat}

In this section we detail the calculation of the negativity ${\cal E}$ introduced 
in Ref.~\onlinecite{ssr-16} for free-fermion systems. 
To define ${\cal E}$, one introduces the fermionic correlation matrix 
$G'_t$ as 
\begin{equation}
G'_{mn}=\delta_{nm}-2G_{nm},\quad n,m=1,\dots,L, 
\end{equation}
where $G_{nm}$ is the same as in Eq.~\eqref{CovMat}. Now, given 
a partition of the system as $A_1\cup A_2$ (in Fig.~\ref{fig0:chain} one 
has $A_1=A$ and $A_2=B$), one defines the matrices $G'_{11},G'_{12},G'_{21},
G'_{22}$ as 
\begin{equation}
\label{eq:gp}
G'=\left(\begin{array}{cc}
	G'_{11} & G'_{12}\\
	G'_{21} & G'_{22}
\end{array}
\right)
\end{equation}
Here $G'_{ij}$ with $i,j=1,2$  are obtained from $G'_{nm}$ by requiring that 
$n\in A_i, m\in A_j$. One now defines the matrices $G^\pm$ as 
\begin{equation}
\label{eq:gpm}
G^\pm=\left(\begin{array}{cc}
	-G'_{11} & \pm iG'_{12}\\
	\pm iG'_{21} & G'_{22}
\end{array}
\right)
\end{equation}
Now one defines the matrix $P$ as 
\begin{equation}
\label{eq:pmatrix}
P\equiv \mathbb{I}+G^+G^-,
\end{equation}
where $\mathbb{I}$ denotes the identiy matrix. 
The key ingredient to define the fermionic negativity 
is the matrix $G^T$ defined as 
\begin{equation}
\label{eq:gt}
G^T\equiv \frac{1}{2}(\mathbb{I}-P^{-1}(G^++G^-)). 
\end{equation}
Given the eigenvalues $\xi_i$ of $G^T$ and the eigenvalues 
$\zeta_i$ of the covariance matrix $G$ (see Eq.~\eqref{CovMat}), 
the fermionic negativity ${\cal E}$ is defined as 
\begin{equation}
\label{eq:negat}
{\cal E}=\sum_j\ln[\xi_j^\frac{1}{2}+(1-\xi_j)^\frac{1}{2}]+
\frac{1}{2}\sum_j\ln[\zeta_j^2+(1-\zeta_j)^2]. 
\end{equation}
%

\subsection{Cancellation of the incoherent contributions}
\label{sec:net-inc}

Here we show that in the presence of gain/loss dissipation the 
incoherent contribution cancels out in the definition of ${\cal E}$, 
at least up to order ${\mathcal O}(b)$, with $b\equiv 
e^{-(\gamma^++\gamma^-)t}$. 

Let us define as $\widetilde G$ as the correlation matrix 
in the absence of dissipation (see the main manuscript). 
In the presence of gain/loss dissipation the 
matrix $G'$ (cf.~\eqref{eq:gp}) is rewritten as  
\begin{equation}
G'=(1-2n_\infty)(1-b)\mathbb{I}+b \widetilde G. 
\end{equation}
Here $n_\infty$ is defined in~\eqref{n-steady}. 
One now has the two matrices $G^\pm$ (cf.~\eqref{eq:gpm}) 
as 
\begin{multline}
\label{eq:gpm-2}
	G^\pm=(1-2n_\infty)\left(\begin{array}{cc}
		-\mathbb{I}_1 & 0\\
		0 &\mathbb{I}_2
	\end{array}\right)\\
-b(1-2a)
\left(\begin{array}{cc}
		-\mathbb{I}_1 & 0\\
		0 &\mathbb{I}_2
	\end{array}\right)
	+b \widetilde G^\pm. 
\end{multline}
Here $\widetilde G^\pm$ are obtained from~\eqref{eq:gpm} after 
replacing $G'_{ij}\to\widetilde G_{ij}$, with $i,j=1,2$. 
Here $\mathbb{I}_{i}$ is the identity matrix restricted to 
subysystem $A_i$. 
It is convenient to separate the different powers of $b$. 
At the leading order in the limit $b\to 0$, i.e., $t\to\infty$,  
after using~\eqref{eq:gpm-2} one has that 
the matrix $P$ (cf.~\eqref{eq:pmatrix}) is diagonal and it is given as 
\begin{equation}
P=(1+(1-2n_\infty)^2)\mathbb{I}. 
\end{equation}
The correlation matrix $G^T$ (cf.~\eqref{eq:gt}) is also diagonal, 
and it is given as 
\begin{equation}
\label{eq:gt-2}
G^T=\frac{1}{2}\left(\begin{array}{cc}
		(1+\frac{2(1-2n_\infty)}{1+(1-2n_\infty)^2})\mathbb{I}_{1} & 0\\
		0 & (1-\frac{2(1-2n_\infty)}{1+(1-2n_\infty)^2})\mathbb{I}_{2}
	\end{array}\right)
\end{equation}
At the leading order, the covariance matrix $G$ (see~\eqref{CovMat}) 
is $n_\infty\mathbb{I}$. 
By using~\eqref{eq:negat}, it is straightforward to show that 
the negativity vanishes in the steady state. 

A similar cancellation occurs for the terms ${\mathcal O}(b)$. 
Up to first order ${\mathcal O}(b)$, $P$ reads 
\begin{multline}
P=(1+(1-2b)(1-2n_\infty)^2)\mathbb{I}\\
+2b(1-2n_\infty)
\left(
\begin{array}{cc}
\widetilde G_{11} & i \widetilde G_{12}\\
-i \widetilde G_{21} & \widetilde G_{22}
\end{array}
\right)
\end{multline}
After keeping only the terms ${\mathcal O}(b)$, 
its inverse $P^{-1}$ reads 
\begin{multline}
	P^{-1}=\frac{1}{1+(1-2n_\infty)^2}\Big[
		\Big(1+\frac{2b(1-2n_\infty)^2}{1+(1-2n_\infty)^2}\Big)\mathbb{I}
		\\
	-\frac{2b(1-2n_\infty)}{1+(1-2n_\infty)^2}
\left(
\begin{array}{cc}
\widetilde G_{11} & i \widetilde G_{12}\\
-i \widetilde G_{21} & \widetilde G_{22}
\end{array}
\right)
\Big]
\end{multline}
%
%
\begin{figure*}[t]
\includegraphics[width=0.8\textwidth]{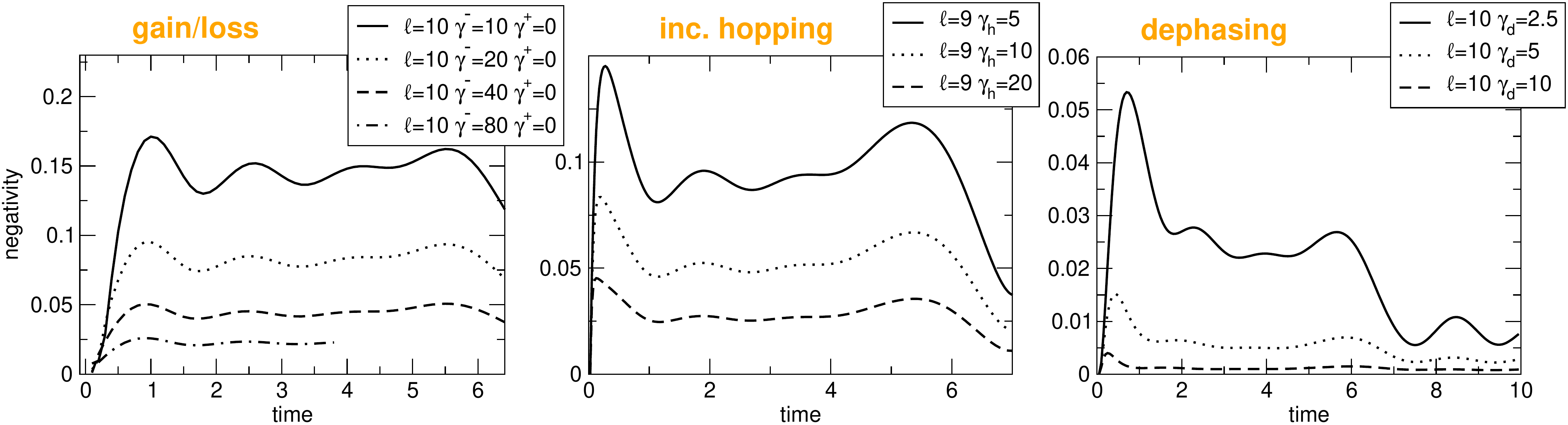}
\caption{ Dynamics of the fermionic negativity ${\cal E}$ in the XX chain 
 in the presence of dissipation: pump/loss of particles (a), 
 incoherent hopping (b), and dephasing (c). To measure the effect of the 
 breaking of the pair structure, the dissipation is present only 
 in the $B$ part of the chain (see Fig.~\ref{fig0:chain}). 
 Note that the negativity plateaux decay as $1/\gamma^-$ and 
 $1/\gamma_h$, whereas for the dephasing the decay is faster as 
 $1/\gamma_d^2$. 
}
\label{fig:broken_pair}
\end{figure*}
%
Up to terms ${\mathcal O}(b)$, the matrix $G^T$ reads 
\begin{multline}
\label{eq:GT}
G^T=\frac{1}{2}\mathbb{I}-\frac{1}{2+2(1-2n_\infty)^2}\Big[
		2(1-2n_\infty)\left(\begin{array}{cc}
			-\mathbb{I}_1 & 0\\
			0 & \mathbb{I}_2
		\end{array}
	\right)
	\\-2b(1-2n_\infty)\left(
	\begin{array}{cc}
	-\mathbb{I}_1 & 0\\
	0 & \mathbb{I}_2
	\end{array}
	\right)
	+b \widetilde G^++\widetilde b G^-
\\+\frac{4b(1-2n_\infty)^3}{1+(1-2n_\infty)^2}
\left(
\begin{array}{cc}
-\mathbb{I}_1 & 0\\
0 & \mathbb{I}_2
\end{array}
\right)
-\frac{4b(1-2n_\infty)^2}{1+(1-2n_\infty)^2}
\widetilde G^+
\Big].
\end{multline}
We now use that the covariance matrix $G$ is  
\begin{equation}
	\label{eq:G}
G=n_\infty (1-b)\mathbb{I}+\frac{b}{2}(\mathbb{I}-\widetilde G)
\end{equation}
To proceed, one has to subsystitute~\eqref{eq:G} and~\eqref{eq:GT} in~\eqref{eq:negat}, 
keeping only the terms ${\cal O}(b)$.  
A straightforward, although tedious, calculation gives that at ${\mathcal O}(b)$ the 
negativity vanishes.

\section{Effect of the pair destruction}
\label{sec:broken-pairs}

In this section we provide some further evidence for the validity of the 
quasiparticles picture for the negativity for the different sources of 
dissipation analyzed in the manuscript. The idea is to consider an 
inhomogeneous dissipation. Specifically, here we consider the case in 
which the Lindbladian acts only on the $B$ part of the chain (see Fig.~\ref{fig0:chain}). 
Moreover, we consider the quench from an inhomogeneous initial state. Specifically, 
we prepare subsystem $A$ in the N\'eel state, whereas part $B$ is prepared in 
the ferromagnetic state. The physical picture is the following. Genuine quantum 
correlations, which are due to the creation of entangled pairs,  are  
generated in $A$. As the quasiparticles travel,  they entangle $A$ and $B$. However, 
the presence of the Lindbladian on part $B$ is expected to suppress the 
negativity, because it destroys one member of the entangled pair, 
implying that the entanglement decays. 

This is explicitly checked in Fig.~\ref{fig:broken_pair} for the 
case of diagonal dissipation (a), incoherent hopping (b), and 
dephasing (c). The figure shows ${\cal E}$ for a chain of 
$L=20$ sites. Subsystem $A$ is the half-chain prepared in 
the ferromagnetic state. For each value of $\gamma^\pm$, 
$\gamma_\mathrm{h}$ and $\gamma_\mathrm{d}$ the negativity 
exhibits a plateaux at $t\ll\ell$. The height of the plateaux 
decays in the limit of large dissipation.

\section{Trace formulas for free fermion quenches}
\label{sec:mau}

In this section we review the calculation of 
arbitrary traces of the correlation matrix after a quench in 
free-fermion models. We follow closely Ref.~\onlinecite{fagotti-2012}. 

First, given the free-fermion creation and annihilation operators 
$c^\dagger_j,c_j$ acting on site $j$, let as define the Majorana 
operators $a_j^x,a_j^y$ as 
\begin{equation}
a_j^x\equiv c_j^\dagger+c_j,\qquad a_j^y\equiv i(c_j-c_j^\dagger).
\end{equation}
Let us define the matrix $\Gamma$ as 
\begin{equation}
	\Gamma_{nm}=
	\left[
		\begin{array}{cc}\delta_{mn}-\langle a_n^xa_m^x\rangle & -\langle a_n^xa_m^y\rangle\\
			-\langle a_n^ya_m^x\rangle & \delta_{mn}-\langle a_n^ya_m^y\rangle
\end{array}\right]. 
\end{equation}
The matrix $\Gamma$ is antisymmetric and has eigenvalues 
$\pm\nu_j$. This are related to the eigenvalues $\lambda_j$ of the fermionic 
correlation matrix $G_{nm}\equiv\langle c^\dagger_n c_m\rangle$ as 
\begin{equation}
\nu_j=2\lambda_j-1.
\end{equation}
The matrix $\Gamma$ is a $2$ by $2$ block Toeplitz matrix as 
\begin{equation}
\Gamma = \left[
 \begin{array}{ccccc}
\mathtt\Gamma_{0}  & \mathtt\Gamma_{-1}   &   \cdots & \mathtt\Gamma_{1-\ell}  \\
\mathtt\Gamma_{1} & \mathtt\Gamma_{0}   & &\vdots\\
\vdots&  & \ddots&\vdots  \\
\mathtt\Gamma_{\ell-1}& \cdots  & \cdots  &\mathtt\Gamma_{0}
\end{array}
\right], ~~~ 
\mathtt\Gamma_{l}=\left(
\begin{array}{cc}
-f_{l}&g_{l}\\
-g_{-l}&f_{l}
\end{array}
\right)\,,
\end{equation}
with $f_l,g_l$ some functions. 
The so-called block symbol $\hat \Gamma(k)$ is defined as the Fourier 
transform 
\begin{multline}
\Gamma_{l}=\left(
\begin{array}{cc}
-f_{l}&g_{l}\\
-g_{-l}&f_{l}
\end{array}
\right)=
\int_{-\pi}^\pi\frac{\mathrm d k}{2\pi} e^{i l k} \hat\Gamma(k)
\,,\quad\\
{\rm with}\quad  \hat\Gamma(k)=
\left(
\begin{array}{cc}
-f(k)&g(k)\\
-g(-k)&f(k)
\end{array}
\right),
\end{multline}
Let us assume that $\hat\Gamma(k)$ can be parametrized as 
\begin{multline}
	\label{tk}
	\hat \Gamma(k)=n_x(k)\sigma^{(k)}_x+\vec n_\perp(k)\cdot\vec\sigma^{(k)}
	e^{2\epsilon(k)t\sigma^{(k)}_x}, \\
	\mathrm{with}\qquad \vec n_\perp(k)\cdot\hat x=0. 
\end{multline}
Here $\sigma_{x,y,z}$ are Pauli matrices, $n_x\in\mathbb{R}$, $t$ a real parameter, 
and $\vec n_\perp$ an arbitrary three-dimensional vector. In~\eqref{tk} 
$\epsilon(k)$ is an arbitrary function, although later we will identify it 
with the single-particle energy of the free-fermion Hamiltonian. 
We also define $\sigma_\alpha^{(k)}$ as 
\begin{equation}
	\sigma_\alpha^{(k)}\sim e^{i\vec w(k)\cdot\vec\sigma}\sigma_\alpha 
	e^{-i\vec w(k)\cdot\vec\sigma},
\end{equation}
and the vector $\vec w(k)$ can be arbitrary. 
The main result is that 
\begin{multline}
\label{mau_formula}
\lim\limits_{t,\ell\to\infty\atop
t/\ell\, \textrm{fixed}}\frac{\textrm{Tr}[\Gamma^{2\beta}]}{2\ell}=\\
\int_{-\pi}^\pi
\frac{dk_0}{2\pi}\textrm{max}\Big(1-2|\epsilon'(k_0)|\frac{t}{\ell},0\Big)\Big(
n_x(k_0)^2+|\vec n_\perp(k_0)|^2\Big)^{2\beta}+\\
\int_{-\pi}^\pi\frac{dk_0}{2\pi}\textrm{min}\Big(2|\epsilon'(k_0)|\frac{t}{\ell},1\Big)
n_x(k_0)^{2\beta}. 
\end{multline}
Here $\epsilon'(k_0)$ is the derivative of $\epsilon$. 
To use~\eqref{mau_formula} we notice that given the symbol 
of the initial correlation 
matrices $\hat\Gamma_0$, at generic time $t$ one has  
\begin{equation}
\hat\Gamma_t(k)=e^{-ih(k)t}\hat\Gamma_0 e^{ih(k)t}, 
\end{equation}
where $h(k)$ is the symbol of the free-fermion Hamiltonian 
generating the dynamics.

Importantly, the results above are valid for 
quenches from translationally invariant initial states. 
Here we are interested in the quench from the N\'eel state, which is 
not translational invariant. However, it is possible to 
restore translational invariance in the initial 
state by perfroming the simple unitary transformation $U$
\begin{equation}
\label{eq:U}
U=\prod_{\mathrm{even}\,j}\sigma_{x,j}. 
\end{equation}
Note that $U$ is product of local unitary transformations, 
which implies that it does not affect the entanglement 
properties of the system. Clearly, $U$ maps the N\'eel state 
in the ferromagnet $|\mathrm{F}\rangle$. 

Here we are interested in quenches in the $XX$ chain, which 
is defined by the Hamiltonian 
\begin{equation}
\label{eq:xx-ham}
H=\sum_j(\sigma_{x,j}\sigma_{x,j+1}+\sigma_{y,j}\sigma_{y,j+1}). 
\end{equation}
Under application of $U$, the XX chain is mapped to 
\begin{equation}
\label{eq:xx-ham-2}
H=\sum_j (\sigma_{x,j}\sigma_{x,j+1}-\sigma_{y,j}\sigma_{y,j+1}).
\end{equation}
This corresponds to the $XY$ chain in the limit of infinite anisotropy. 
Eq.~\eqref{eq:xx-ham} is diagoanlized by a Fourier transform. One obtains the 
single particle dispersion as 
\begin{equation}
\epsilon(k)=\sin k. 
\end{equation}
%
\begin{figure}[t]
\includegraphics[width=0.4\textwidth]{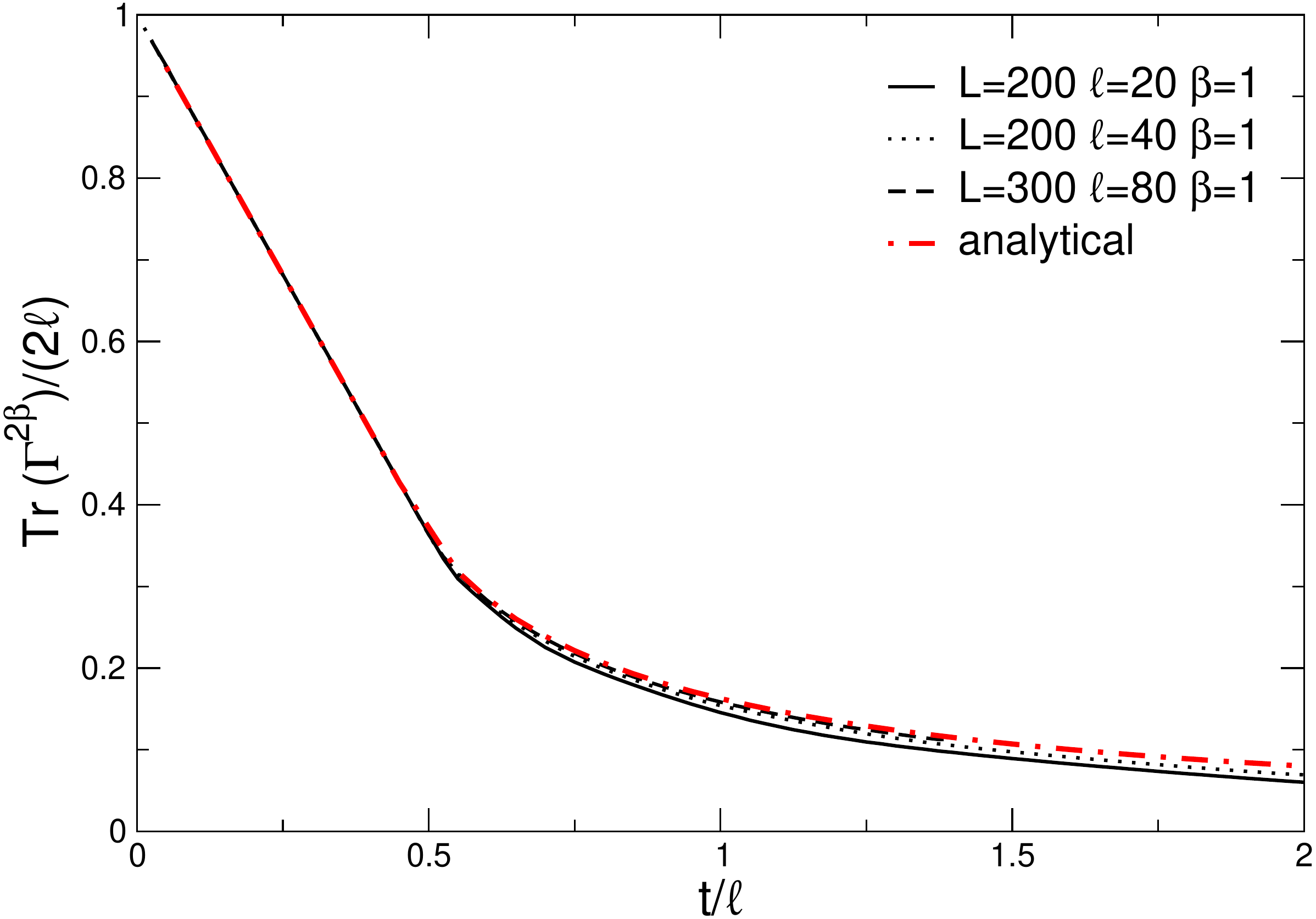}
\caption{ Check of formula~\eqref{mau-formula} for the $XX$ chain and 
 the quench from the N\'eel state. 
}
\label{mau_check}
\end{figure}
%
The symbol $h(k)$ of the Hamiltonian above is 
\begin{equation}
h(k)=\sin k\sigma_x. 
\end{equation}
Before applying~\eqref{mau_formula} we need to determine the 
evolved symbol $\hat \Gamma$ of the correlation matrix. 
For the ferromagnet $|\mathrm{F}\rangle$, a straightforward calculation gives 
\begin{equation}
\Gamma=\delta_{nm}\sigma_y. 
\end{equation}
Now the the symbol of the evolved correlation matrix $\hat \Gamma_t$ at time $t$ 
reads as 
\begin{equation}
\label{eq:tk-1}
\hat\Gamma_t\equiv e^{-i h(k)t}\sigma_y e^{ih(k)t}=e^{-i \sin(k)\sigma_x}\sigma_y e^{i\sin(k)\sigma_x}.
\end{equation}
This is of the form in Eq.~\eqref{tk} with 
\begin{equation}
\label{eq:supp}
\vec w=(\sin(k),0,0),\quad n_x=0,\quad \vec n_\perp=(0,1,0). 
\end{equation}
It is now straightforward to apply~\eqref{mau_formula}. 
The integration over $k_0$ can be performed analytically, and 
one obtains that 
\begin{multline}
\label{mau-formula}
\frac{\textrm{Tr}(\Gamma^{2\beta})}{2\ell}=\Big(1-\frac{4}{\pi}\Big)\frac{t}{\ell}\theta(\ell-2t)
\\+\frac{2}{\pi}\Big[-2\frac{t}{\ell}+\sqrt{\frac{4t^2}{\ell^2}-1}
+\textrm{arccsc}\big(\frac{2t}{\ell}\big)\Big]\theta(2t-\ell). 
\end{multline}
The result does not depend on the index $\beta$. This is a feature of the N\'eel 
state, and it can be derived by observing that the overlaps between the 
N\'eel state and the eigenstates of the $XX$ chain are all equal~\cite{mazza-2016}. 

In Fig.~\ref{mau_check} we check the validity of Eq.~\eqref{mau_formula}. 
Clearly, although scaling corrections are present, upon increasing 
$t,\ell$, one recovers~\eqref{mau_formula}. It is now clear that by 
using the large $t$ expansion in~\eqref{boxed} and~\eqref{mau-formula}, it is 
straightforward to prove~\eqref{boxed-1}. Finally, a similar calculation 
yields the coherent contribution to the R\'enyi entropies 
$S^{\scriptscriptstyle(\alpha)}_\mathrm{q}$ as 
\begin{multline}
\label{renyi-theo}
S_\mathrm{q}^{(\alpha)}
\approx
\frac{-\alpha t e^{-2(\gamma^++\gamma^-)t}}
	{2\pi(1-\alpha)n_\infty^2(n_\infty-1)^2((1-n_\infty)^\alpha+n_\infty^\alpha)^2}\times\\
		((1-n_\infty)^\alpha n_\infty+(n_\infty-1)n_\infty^\alpha)^2
		-n_\infty^\alpha(1-n_\infty)^\alpha (\alpha-1)
\end{multline}
%

\subsection{Resummation formula}
\label{sec:resum}

Formula~\eqref{mau_formula} allows to obtain the behavior of arbitrary functions 
of $\Gamma$ in the scaling limit. Precisely, let us consider 
\begin{equation}
\label{eq:fn}
\mathrm{Tr}{\mathcal 
F}(\Gamma^2)=\sum_\beta{\mathcal F}_\beta\mathrm{Tr}(\Gamma^{2\beta}), 
\end{equation}
where we assumed that ${\mathcal F}(z)$ is an analytic function of 
$z$ with power series expansion ${\mathcal F}=\sum_\beta {\mathcal F}_\beta z^\beta$ 
around $z=0$. Now, one can use~\eqref{mau_formula}, interchange the order 
of summation and integration, to obtain 
\begin{multline}
\label{eq:mau-2}
\lim_{t,\ell\to\infty\atop t/\ell\ {\rm fixed}}
\frac{\mathrm{Tr}[{\cal F}(\Gamma^2)]}{2\ell}=\\
\int_{-\pi}^{\pi}\frac{\mathrm d
  k_0}{2\pi}\max\Big (1-2|\epsilon'(k_0)| \frac{t}\ell,0\Big)  
{\cal F}\bigl(n_x( k_0)^2+|\vec n_\perp( k_0)|^2\bigr)\nonumber\\
\qquad+\int_{-\pi}^\pi \frac{\mathrm d  
k_0}{2\pi}\min\Bigl(2|\epsilon'(k_0)|\frac{t}\ell,1\Bigr){\cal F}(n_x( k_0)^{2}).
\end{multline}
Before applying this result to our problem, it is crucial to observe that 
for generic values of the gain/loss rate $\gamma^\pm$, due to the structure 
of the eigenvalues $\lambda_i$ (cf.~\eqref{eig}), the entropies $S^{\scriptscriptstyle(\alpha)}$ are not even functions of the unitarily evolved eigenvalues $\widetilde 
\nu_j$. Fortunately, this does not happen 
in the balanced gain/loss case, i.e., for 
$\gamma^+=\gamma^-$. Then one has that the function ${\mathcal F}(z)$ is given 
as 
\begin{multline}
	{\mathcal F}(z)=-\frac{1}{2}(1-e^{-(\gamma^++\gamma^-)t}z)
	\ln\frac{1}{2}(1-e^{-(\gamma^++\gamma^-)t}z)\\
-\frac{1}{2}(1+e^{-(\gamma^++\gamma^-)t}z)
	\ln\frac{1}{2}(1+e^{-(\gamma^++\gamma^-)t}z),
\end{multline}
where $\gamma^+=\gamma^-$. 
A straightforward calculation using~\eqref{eq:mau-2} gives~\eqref{main-formula}. 
As it is shown in the main manuscript, although here we proved~\eqref{main-formula} 
for $\gamma^+=\gamma^-$, we checked that it holds for arbitrary $\gamma^\pm$.

\end{appendix}

\end{document}